\documentclass[aip,amsmath,amssymb,reprint,]{revtex4-1}

\usepackage{graphicx}
\usepackage{dcolumn}
\usepackage{bm}
\usepackage{xcolor}
\usepackage[utf8]{inputenc}
\usepackage[T1]{fontenc}
\usepackage{mathptmx}
\usepackage{etoolbox}
\usepackage{bbold}
\usepackage{braket}
\usepackage{dsfont}
\usepackage{tikz}
\usetikzlibrary{quantikz,arrows,positioning,automata,shadows,fit,shapes}
\usepackage{hyperref}
\usepackage{mathdots}
\usepackage{multirow}
\usepackage{tabularx}
\usepackage{soul}

\tikzset{
operator/.append style={fill=gray!5}
}

\makeatletter
\def\@email#1#2{
 \endgroup
 \patchcmd{\titleblock@produce}
  {\frontmatter@RRAPformat}
  {\frontmatter@RRAPformat{\produce@RRAP{*#1\href{mailto:#2}{#2}}}\frontmatter@RRAPformat}
  {}{}
}
\makeatother
\begin{document}

\preprint{AIP/123-QED}

\title[Shallow unitary decompositions of quantum Fredkin and Toffoli gates for connectivity-aware equivalent circuit averaging]{Shallow unitary decompositions of quantum Fredkin and Toffoli gates \\
for connectivity-aware equivalent circuit averaging}
\author{Pedro M. Q. Cruz}
  \email{pedro.cruz@inl.int}
  \affiliation{International Iberian Nanotechnology Laboratory (INL), 4715-330 Braga, Portugal}
  \affiliation{\mbox{ICFO - Institut de Ci\`{e}ncies Fot\`{o}niques, The Barcelona Institute of Science and Technology, 08860 Castelldefels, Spain}}
  \affiliation{\mbox{Departamento de Ci\^encia de Computadores, Universidade do Porto, 4169-007 Porto, Portugal}}

\author{Bruno Murta}
\affiliation{International Iberian Nanotechnology Laboratory (INL), 4715-330 Braga, Portugal}
\affiliation{Departamento de F\'{i}sica, Universidade do Minho, 4710-057 Braga, Portugal}

\date{\today}

\begin{abstract}

The controlled-\textsc{swap} and controlled-controlled-\textsc{not} gates
are at the heart of the original proposal of reversible classical computation by Fredkin and Toffoli. Their widespread use in quantum computation, both in the implementation of classical logic subroutines of quantum algorithms and in quantum schemes with no direct classical counterparts, has made it imperative early on to pursue their efficient decomposition in terms of the lower-level gate sets native to different physical platforms. Here, we add to this body of literature by providing several logically equivalent circuits for the Toffoli and Fredkin gates under all-to-all and linear qubit connectivity, the latter with two different routings for control and target qubits. Besides achieving the lowest \textsc{cnot} counts in the literature for all these configurations, we also demonstrate the remarkable effectiveness of the obtained decompositions at mitigating coherent errors on near-term quantum computers via equivalent circuit averaging. We first quantify the performance of the method in silico with a coherent-noise model before validating it experimentally on a superconducting quantum processor.
In addition, we consider the case where the three qubits on which the Toffoli or Fredkin gates act nontrivially are not adjacent, proposing a novel scheme to reorder them that saves one \textsc{cnot} for every \textsc{swap}. This scheme also finds use in the shallow implementation of long-range \textsc{cnot}s. Our results highlight the importance of considering different entangling gate structures and connectivity constraints when designing efficient quantum circuits.

\end{abstract}

\maketitle

\section{Introduction \label{sec:intro}}

The Fredkin gate (also known as controlled-\textsc{swap}) and the Toffoli gate (also known as controlled-controlled-\textsc{not}) are three-input, three-output logic gates that were introduced within the reversible logic model of classical computation\cite{fredkin1982conservative}, in which logic circuits realize invertible Boolean functions \footnote{In other words, circuits that implement a bijection between input and output states. When, additionally, the circuit outputs a bit string that is just a permutation of the input bits, the computation is said to be \emph{conservative}.}. The \textsc{cswap} leaves an input bit unchanged and swaps the remaining two if and only if the first one is in state 1 \footnote{In fact, in Fredkin and Toffoli\textquotesingle s seminal paper \cite{fredkin1982conservative}, the target-bits are assumed to be swapped only when the control-bit is 0 (see truth table in Eq.~(2)). However, since this paper is devoted to the quantum versions of the Fredkin and Toffoli gates, and because it is is common practice in quantum information science to associate the nontrivial action of a controlled-gate with the control-qubit in state $\ket{1}$, we will adopt this convention.}, while the \textsc{ccnot} negates the target bit if both control bits are in state 1.
Their importance lies in both being universal Boolean primitives for reversible logic: any classical logic operation can be constructed entirely out of Fredkin or Toffoli gates\footnote{Strictly speaking, the classical universality of the Toffoli and Fredkin gates assumes that we can add ancillary bits to the circuit that can be initialized in either 0 or 1 as required.}.

As their reversible nature implies unitarity, both gates were readily adopted in quantum computing, particularly to realize classical logic circuits that perform subroutines of quantum algorithms. As a result, they gain the ability  to operate over superposition states --- i.e., complex-valued linear combinations of the classical states --- and implement arbitrary classical logic operations on quantum data. Moreover, adding just the Hadamard gate to the Toffoli gate suffices to form a universal quantum basis set \cite{shi2002both, aharonov2003asimple}. The Fredkin gate requires the \textsc{x} gate in addition to the Hadamard gate to form a universal quantum basis\footnote{Classically, given the universality of the Fredkin gate, it is clear that a Toffoli gate can be decomposed in terms of Fredkin gates alone by adding extra bits, so it might seem that the basis set comprising the Fredkin and Hadamard is also universal. However, because the Fredkin gate is conservative (i.e., it preserves the Hamming weight of input bitstrings, unlike the Toffoli), this replacement of every Toffoli gate by Fredkin gates would not reset all required ancillary qubits, thus generating undesired garbage qubits. Hence, the \textsc{not} gate needs to be added to the Fredkin and Hadamard gates to form a universal basis set.}. In practice, however, two-qubit gates are often used instead of the Toffoli or Fredkin gates as the elements of basis gate sets that can change the entanglement structure of the input state, a necessary condition for quantum universality. 

Nevertheless, the Fredkin and Toffoli gates play a pivotal role in quantum computation. In particular, the Toffoli gate is the key building block of its multi-qubit generalizations\cite{barenco1995gates, nielsen2002quantum, shende2009on}, which are ubiquitous in quantum arithmetic circuits \cite{vedral1996quantum} and in the construction of oracles \cite{Grover1996, gilliam2020oracles}. Moreover, the Toffoli gate has been adopted in quantum error correction \cite{reed2012realization}. Recently, the iToffoli gate, a close variant of the Toffoli gate\footnote{The iToffoli gate amounts to a Toffoli gate that triggers the \textsc{not} gate on the target-qubit when the control-qubits are in state $|0\rangle$, followed by a controlled-\textsc{s} gate also triggered by $|0\rangle$. See Fig. 1(c) of Ref.\cite{kim2022high}.}, was part of a proposal to compute frequency-domain molecular response properties\cite{sun2023quantum}. As for the Fredkin gate, it is the core element of the \textsc{swap} test \cite{barenco1997stabilization, buhrman2001quantum}, the canonical method to compute the fidelity between two states. In addition, the Fredkin gate has also been employed in quantum state preparation \cite{ozaydin2014fusing, araujo2021divide, murta2023preparing}, estimation of linear and nonlinear functionals of density operators \cite{ekert2002direct}, quantum switches \cite{chiribella2013quantum, araujo2017quantum, castroruiz2018dynamics}, optimal quantum cloning \cite{hofmann2012how}, stabilization of quantum computations by symmetrization\cite{barenco1997stabilization}, sampling states in the Hamiltonian eigenbasis (along with the Toffoli gate) \cite{babbush2018encoding}, and calculation of Bargmann invariants\cite{oszmaniec2021measuring, wagner2023quantum}. Both the Fredkin and Toffoli gates have found use in routines tailored to near-term quantum hardware\cite{murta2023preparing, wagner2023quantum, tacchino2019an, satoh2020subdivided, duckering2021orchestrated, murta2021gutzwiller}.

In light of such a broad range of applications, it is unsurprising that the problem of implementing the Fredkin and Toffoli gates on digital quantum computers has attracted great interest. Unlike previous proposals tailored to specific quantum hardware --- e.g., in platforms based on trapped ions \cite{ivanov2015efficient, rasmussen2020singlestep}, superconducting circuits \cite{fedorov2012use, reed2012realization, zahedinejad2015high, kim2022high, bowman2022hardware} and quantum optics \cite{milburn1989quantum, fiurasek2006linearoptics, gong2008methods, patel2016aquantum, li2022quantum} --- we follow a high-level, hardware-agnostic approach, whereby the Fredkin and Toffoli gates are decomposed in terms of standard single- and two-qubit operations. In particular, we take the \textsc{cnot} as the reference two-qubit basis gate. Earlier works\cite{amy2013a, jones2013lowoverhead, selinger2013quantum} have minimized the number of non-Clifford operations such as \textsc{t} gates to render the Fredkin and Toffoli less onerous for fault-tolerant quantum computation\cite{preskill1997fault}. Instead, we focus on decompositions suitable for noisy intermediate-scale quantum hardware\cite{preskill2018quantum}, in which case the key goal is to minimize the number of \textsc{cnot} gates whilst taking qubit connectivity constraints into account. A lower \textsc{cnot} count can be achieved by allowing for implementations up to a relative phase factor\cite{barenco1995gates, selinger2013quantum, maslov2016advantages} or by replacing some qubits with qutrits\cite{ralph2007efficient, baekkgaard2019realization, gokhale2019asymptotic, liu2020optimal}. Here, we restrict ourselves to the consideration of qubits, aiming to realize three-qubit operations with the exact matrix representations shown in Fig.~\ref{fig:standard-decomps}, up to a global phase.
 
The remainder of the paper is structured as follows. Section \ref{sec:adj-dec} considers the \textsc{cnot}-count minimization of the Fredkin and Toffoli gate decompositions for three adjacent qubits with both all-to-all and linear qubit connectivity. 
Section \ref{sec:nonadj-dec} contemplates the case where the three qubits on which these unitaries act nontrivially are not directly connected to one another. In particular, we devise a method to bring the three qubits together and then return them to their original positions that saves one \textsc{cnot} for every \textsc{swap}. In Section \ref{sec:ECA}, we exploit the multiple generated circuits for the Fredkin and Toffoli gates to mitigate coherent errors via equivalent circuit averaging, analyzing performance in silico and experimentally. Lastly, Section \ref{sec:concl} summarizes our results.

\section{Decompositions for adjacent qubits \label{sec:adj-dec}}

\begin{figure}[t]
\begin{raggedright}
\begin{tabular}{lll}
(a) \, \raisebox{-0.8\totalheight}{
\includegraphics[width=0.39\linewidth]{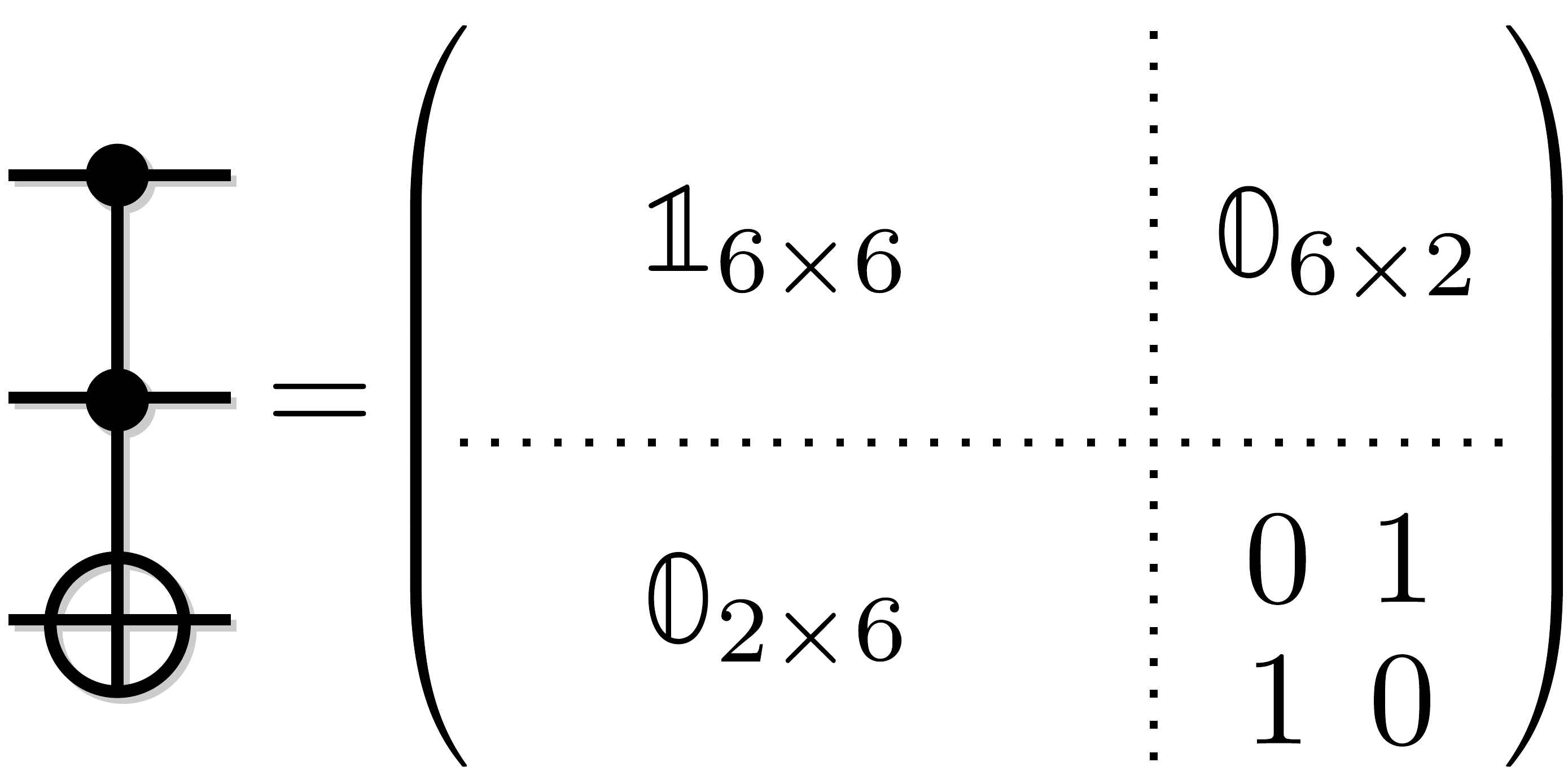}
} &  & (b) \, \raisebox{-0.8\totalheight}{
\includegraphics[width=0.39\linewidth]{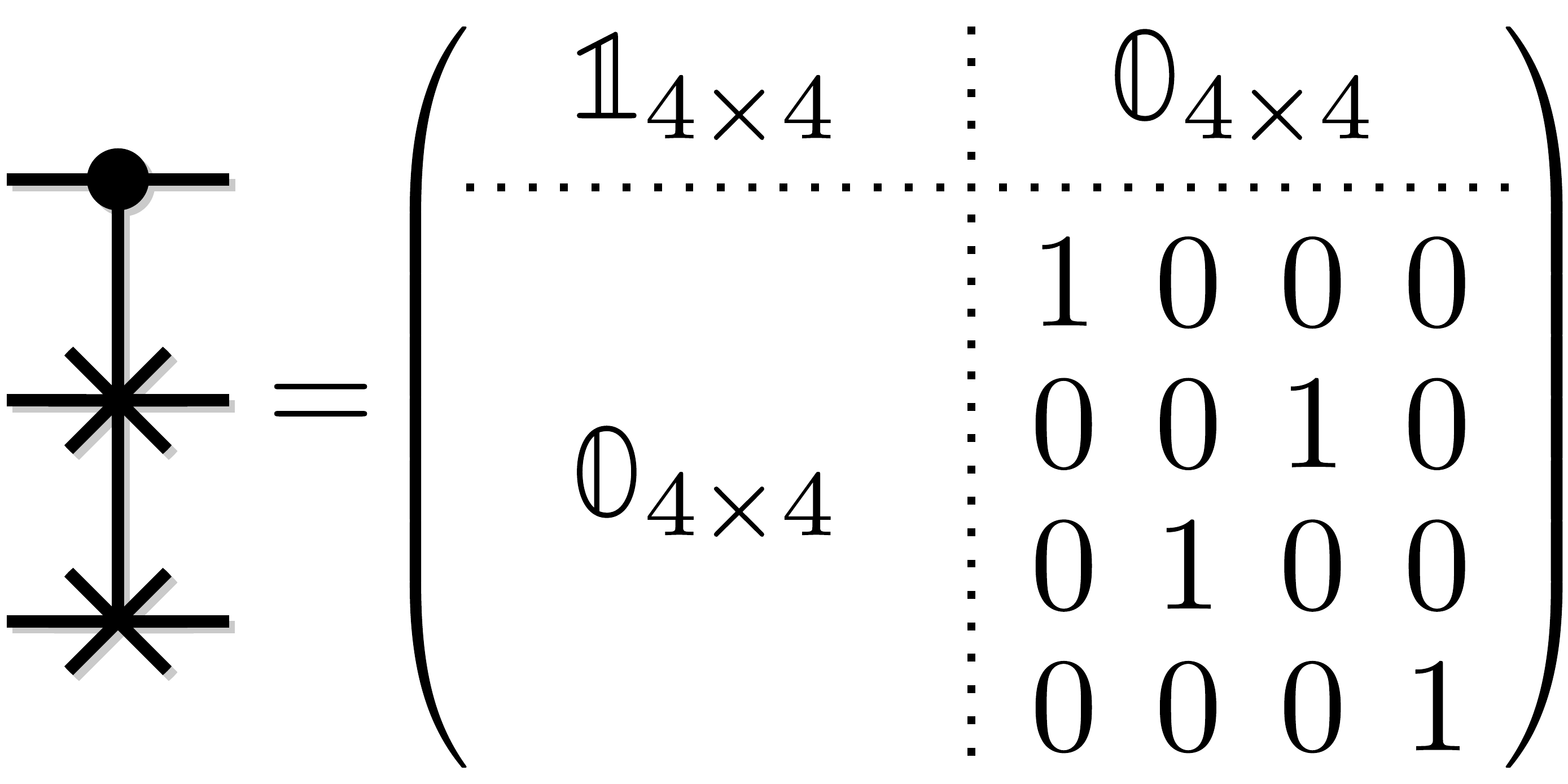}
}
\tabularnewline
\tabularnewline
(c)\raisebox{-0.7\totalheight}{
\begin{tikzpicture}
\node[scale=0.8]{
\begin{tikzcd}[row sep={0.5cm,between origins}, column sep=0.18cm]
& \swap{1} & \qw \midstick[2,brackets=none]{$=$} & \targ{} & \ctrl{1} & \targ{} & \qw \\
& \targX{} & \qw & \ctrl{-1} & \targ{} & \ctrl{-1} & \qw \\
\end{tikzcd}
};
\end{tikzpicture}
} &  & (d)\raisebox{-0.7\totalheight}{
\begin{tikzpicture}
\node[scale=0.8]{
\begin{tikzcd}[row sep={0.5cm,between origins}, column sep=0.18cm]
& \ctrl{2} & \qw \midstick[3,brackets=none]{$=$} & \qw & \ctrl{2} & \qw & \qw \\
& \swap{1} & \qw & \targ{} & \ctrl{1} & \targ{} & \qw \\
& \targX{} & \qw & \ctrl{-1} & \targ{} & \ctrl{-1} & \qw
\end{tikzcd}
};
\end{tikzpicture}
}\tabularnewline
\multicolumn{3}{l}{(e)\raisebox{-0.7\totalheight}{
\begin{tikzpicture}
\node[scale=0.85]{
\begin{tikzcd}[row sep={0.7cm,between origins}, column sep=0.2cm]
& \qw & \qw \gategroup[3,steps=12,style={solid, rounded corners, draw=blue, inner xsep=0pt, inner ysep=0pt}]{} & \qw & \qw & \ctrl{2} & \qw & \qw & \qw & \ctrl{2} & \ctrl{1} & \qw & \ctrl{1} & \gate{T} & \qw & \qw \\
& \targ{} \gategroup[2,steps=4,style={dashed, rounded corners, draw=red, inner xsep=0pt, inner ysep=3pt}]{} & \qw & \ctrl{1} & \qw & \qw & \qw & \ctrl{1} & \gate{T^{\dagger}} & \qw & \targ{} & \gate{T^{\dagger}} & \targ{} & \gate{S} & \targ{} & \qw \\
& \ctrl{-1} & \gate{H} & \targ{} & \gate{T^{\dagger}} & \targ{} & \gate{T} & \targ{} & \gate{T^{\dagger}} & \targ{} & \gate{T} & \gate{H} & \qw & \qw & \ctrl{-1} & \qw
\end{tikzcd}
};
\end{tikzpicture}
}}\tabularnewline
\multicolumn{3}{l}{(f)\raisebox{-0.7\totalheight}{
\begin{tikzpicture}
\node[scale=0.85]{
\begin{tikzcd}[row sep={0.7cm,between origins}, column sep=0.2cm]
& \qw & \qw & \qw & \qw & \ctrl{2} & \qw & \qw & \qw & \ctrl{2} & \ctrl{1} & \gate{T} & \ctrl{1} & \qw & \qw \\
& \gate{S} \gategroup[2,steps=4,style={dashed, rounded corners, draw=red, inner xsep=0pt, inner ysep=0pt}]{} & \targ{} & \gate{S^{\dagger}} & \qw & \qw & \qw & \ctrl{1} & \gate{T} & \qw & \targ{} & \gate{T^{\dagger}} & \targ{} & \targ{} & \qw \\
& \qw & \ctrl{-1} & \gate{\sqrt{X}} & \gate{T} & \targ{} & \gate{T} & \targ{} & \gate{T^{\dagger}} & \targ{} & \gate{T} & \gate{H} & \qw & \ctrl{-1} & \qw
\end{tikzcd}
};
\end{tikzpicture}
}}\tabularnewline
\end{tabular}
\par\end{raggedright}
\caption{
Matrix representations and quantum circuit diagrams of (a) the Fredkin gate, and (b) the Toffoli gate. Qubit significance decreases from top to bottom in the diagrams.
(c) Standard decomposition of \textsc{swap} gate in terms of $3$ \textsc{cnot}s. (d) Decomposition of Fredkin gate in terms of a Toffoli gate between two \textsc{cnot}s, adapting the circuit in (c) and making use of the result from Appendix \ref{AppA}. (e) Introduction of textbook decomposition of Toffoli gate \cite{nielsen2002quantum}, as depicted within solid-line blue box, into circuit shown in (d). This gives rise to a circuit for the Fredkin gate with $8$ \textsc{cnot}s and depth $14$. (f) Two simplifications to the circuit shown in (e) can be applied. The first corresponds to the replacement of the two-qubit subcircuit shown inside the dashed-line red box, which results in one less \textsc{cnot}. The second amounts to removing a layer of single-qubit gates by changing some single-qubit gates whilst keeping the \textsc{cnot} structure unaltered. Overall, the Fredkin gate on three adjacent qubits can therefore be executed with $7$ \textsc{cnot}s and a depth of $13$, ignoring qubit connectivity constraints. Single-qubit Hadamard ($H$), phase ($S$) and $\pi/8$ ($T$) gates follow the standard definitions\cite{nielsen2002quantum}, and $\sqrt{X} = H S H$.}
\label{fig:standard-decomps}
\end{figure}
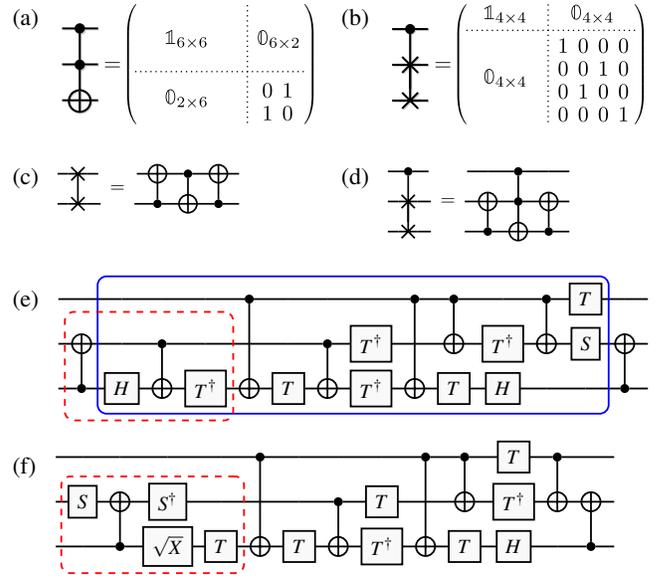

It is well established that five two-qubit operations suffice to decompose the Toffoli gate\cite{smolin1996five, yu2013optimal}. However, the native basis gate sets that can be realized in quantum computing platforms typically only include a single fixed (i.e., not parameterized) two-qubit operation such as the \textsc{cnot}. Hence, in practice, the minimum number of two-qubit gates involved in the decomposition of the Toffoli gate is $6$. The circuit inside the blue solid-line box in Fig.~\ref{fig:standard-decomps}(e) shows the textbook decomposition\cite{nielsen2002quantum} of the Toffoli gate, which is optimal as far as the \textsc{cnot} count is concerned. Henceforth, this circuit will be the starting point to find shallow decompositions of the Toffoli and Fredkin gates under different qubit connectivity constraints.

The standard quantum circuit for the Fredkin gate \cite{smolin1996five} results from adapting the well-known decomposition of the \textsc{swap} gate in terms of  $3$ \textsc{cnot}s (Fig.~\ref{fig:standard-decomps}(c)). Na\"{i}vely, an extra control-qubit should be added to each \textsc{cnot}, but only the middle one happens to be required (Fig.~\ref{fig:standard-decomps}(d)) thanks to the symmetric structure of the \textsc{swap} circuit (see Appendix~\ref{AppA}). Making use of the aforementioned textbook decomposition of the Toffoli gate \cite{nielsen2002quantum}, this results in a circuit for the Fredkin gate with $8$ \textsc{cnot}s and depth $14$ (Fig.~\ref{fig:standard-decomps}(e)). However, the subcircuit within the red dashed-line box can be further simplified, resulting in the elimination of $1$ \textsc{cnot}. Moreover, a layer of single-qubit gates can also be removed at the end of the circuit by changing some single-qubit gates whilst leaving the entangling gates structure unchanged. The result of these two simplifications is shown in Fig.~\ref{fig:standard-decomps}(f), corresponding to a total of $7$ \textsc{cnot}s and a circuit depth of $13$. To the best of our knowledge, this is the shallowest decomposition of the Fredkin gate in the literature in terms of \textsc{cnot} count.

The circuits shown in Fig.~\ref{fig:standard-decomps} assume all qubits are connected to one another, thus allowing to implement a \textsc{cnot} gate between any pair of qubits natively. However, in quantum computers based on solid-state platforms that realize qubits through superconducting circuits\cite{blais2021circuit} or silicon quantum dots\cite{zwanenburg2013silicon}, there are unavoidable restrictions in the connections between qubits. \textsc{cnot} gates between widely separated qubits are only possible by moving the information content of the qubits around through networks of $\textsc{swap}$ gates \cite{ogorman2019generalized}, which introduce a considerable depth overhead. Generating shallow decompositions that forgo such \textsc{swap} networks whilst taking these qubit connectivity constraints into account is thus crucial to exploit the potential of near-term quantum processors. This is particularly relevant for circuits comprising three-qubit operations such as the Toffoli or the Fredkin gates, as the architectures of most quantum processors that are currently available or under development do not include trios of fully connected qubits.

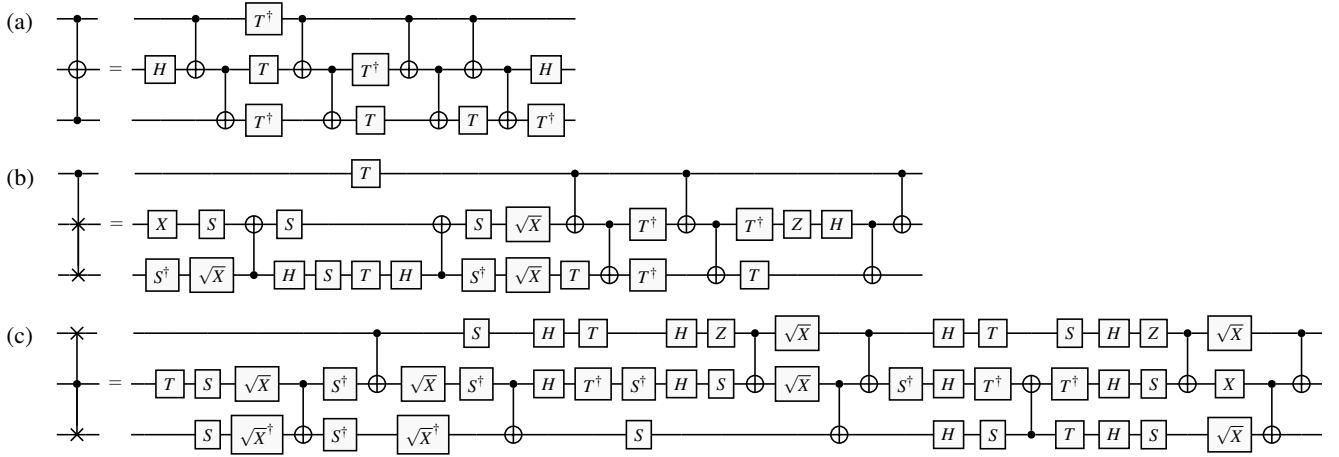
\begin{figure*}[t]
\begin{raggedright}
\begin{tabular}{l}
(a) \raisebox{-0.76\totalheight}{
\begin{tikzpicture}
\node[scale=0.75]{
\begin{tikzcd}[row sep={0.9cm,between origins}, column sep=0.2cm]
& \ctrl{1} & \qw \midstick[3,brackets=none]{$=$} & \qw & \ctrl{1} & \qw & \gate{T^{\dagger}} & \ctrl{1} & \qw & \qw & \ctrl{1} & \qw & \ctrl{1} & \qw & \qw & \qw \\
& \targ{} & \qw & \gate{H} & \targ{} & \ctrl{1} & \gate{T} & \targ{} & \ctrl{1} & \gate{T^{\dagger}} & \targ{} & \ctrl{1} & \targ{} & \ctrl{1} & \gate{H} & \qw \\
& \ctrl{-1} & \qw & \qw & \qw & \targ{} & \gate{T^{\dagger}} & \qw & \targ{} & \gate{T} & \qw & \targ{} & \gate{T} & \targ{} & \gate{T^{\dagger}} & \qw
\end{tikzcd}
};
\end{tikzpicture}
} \\
(b) \raisebox{-0.76\totalheight}{
\begin{tikzpicture}
\node[scale=0.75]{
\begin{tikzcd}[row sep={0.9cm,between origins}, column sep=0.2cm]
& \ctrl{2} & \qw \midstick[3,brackets=none]{$=$} & \qw & \qw & \qw & \qw & \qw & \gate{T} & \qw & \qw & \qw & \qw & \ctrl{1} & \qw & \qw & \ctrl{1} & \qw & \qw & \qw & \qw & \qw & \ctrl{1} & \qw \\
& \targX{} & \qw & \gate{X} & \gate{S} & \targ{} & \gate{S} & \qw & \qw & \qw & \targ{} & \gate{S} & \gate{\sqrt{X}} & \targ{} & \ctrl{1} & \gate{T^{\dagger}} & \targ{} & \ctrl{1} & \gate{T^{\dagger}} & \gate{Z} & \gate{H} & \ctrl{1} & \targ{} & \qw \\
& \swap{-1} & \qw & \gate{S^{\dagger}} & \gate{\sqrt{X}} & \ctrl{-1} & \gate{H} & \gate{S} & \gate{T} & \gate{H} & \ctrl{-1} & \gate{S^{\dagger}} & \gate{\sqrt{X}} & \gate{T} & \targ{} & \gate{T^{\dagger}} & \qw & \targ{} & \gate{T} & \qw & \qw & \targ{} & \qw & \qw
\end{tikzcd}
};
\end{tikzpicture}
} \\
(c) \raisebox{-0.76\totalheight}{
\begin{tikzpicture}
\node[scale=0.75]{
\begin{tikzcd}[row sep={0.9cm,between origins}, column sep=0.2cm]
& \targX{} & \qw \midstick[3,brackets=none]{$=$} & \qw  & \qw & \qw & \qw  & \qw & \qw & \ctrl{1} & \qw & \gate{S} & \qw  & \gate{H} & \gate{T} & \qw & \gate{H} & \gate{Z} & \ctrl{1} & \gate{\sqrt{X}} & \qw & \ctrl{1} & \qw & \gate{H} & \gate{T} & \qw & \gate{S} & \gate{H} & \gate{Z} & \ctrl{1} & \gate{\sqrt{X}} & \qw & \ctrl{1} & \qw \\
& \ctrl{1} & \qw & \qw  & \gate{T}  & \gate{S} & \gate{\sqrt{X}} & \ctrl{1} & \gate{S^{\dagger}} & \targ{} & \gate{\sqrt{X}} & \gate{S^{\dagger}} & \ctrl{1} & \gate{H} & \gate{T^{\dagger}} & \gate{S^{\dagger}} & \gate{H} & \gate{S} & \targ{} & \gate{\sqrt{X}} & \ctrl{1} & \targ{} & \gate{S^{\dagger}} & \gate{H} & \gate{T^{\dagger}} & \targ{} & \gate{T^{\dagger}} & \gate{H} & \gate{S} & \targ{} & \gate{X} & \ctrl{1} & \targ{} & \qw \\
& \swap{-2} & \qw & \qw & \qw & \gate{S} & \gate{\sqrt{X}^{\dagger}} & \targ{} & \gate{S^{\dagger}} & \qw & \gate{\sqrt{X}^{\dagger}} & \qw & \targ{} & \qw & \qw & \gate{S} & \qw & \qw & \qw & \qw & \targ{} & \qw & \qw & \gate{H} & \gate{S} & \ctrl{-1} & \gate{T} & \gate{H} & \gate{S} & \qw & \gate{\sqrt{X}} & \targ{} & \qw & \qw
\end{tikzcd}
};
\end{tikzpicture}
} \\
\end{tabular}
\par\end{raggedright}
\caption{Examples of optimal basis gate decompositions in \textsc{cnot} count obtained through our ZX-calculus-based optimization heuristic for the case of linear qubit connectivity  of (a) Toffoli gate, (b) Fredkin gate with control-qubit at one end of three-qubit register and (c) Fredkin gate with control-qubit at center of three-qubit register. Single-qubit Hadamard ($H$), phase ($S$) and $\pi/8$ ($T$), Pauli-X ($X$) and Pauli-Z ($Z$) gates follow the standard definitions\cite{nielsen2002quantum}, and $\sqrt{X} = H S H$, being $\sqrt{X}^{\dagger}$ its inverse. The circuit in (a) applies specifically to the case where the target-qubit of the Toffoli gate is at the central position, but the target-qubit can be changed by simply moving the two Hadamard gates, one on either end of the circuit, to the desired target-qubit (see Appendix \ref{AppB}).}
\label{fig:lin-decomps}
\end{figure*}

Leveraging the ZX-calculus and optimization heuristics, we have recently developed a technique\cite{cruz2024} for unitary decomposition capable of producing many logically equivalent circuits with manifestly different entangling gate structures. The entangling gate structure of the circuit, as we define it, consists of the description of the order and position of the \textsc{cnot} gates applied to different qubit pairs along the execution of the circuit. Single-qubit gates are excluded from this definition, grouping circuits differing only in single-qubit gates under the same category. Furthermore, if two circuits differ from each other only due to permutations of commuting \textsc{cnot} gates, they are also considered under the same entangling gate structure\cite{cruz2024}.

The input provided to this circuit optimization technique is a circuit that implements the desired gate; this initial circuit is generally suboptimal in \textsc{cnot} count, and the goal of the method is to generate an equivalent circuit with fewer \textsc{cnot}s. However, it is equally possible to start from a \textsc{cnot}-count-optimal circuit and obtain another circuit with the same number of \textsc{cnot}s but a different entangling gate structure.
The input circuit is converted into a ZX-diagram through the \texttt{PyZX} software package\cite{kissinger2020pyzx}, which also includes methods to simplify the ZX-diagram and convert it back into a quantum circuit\cite{duncan2020graphtheoretic, backens2021there}. This conversion can often give rise to a wide variety of circuits, and our technique searches for those that minimize the \textsc{cnot} count. Specifically, we build upon the \texttt{PyZX} simplification techniques with an intensive search and optimization procedure that often succeeds in escaping from local minima, thus optimizing the decompositions further.

We have applied our circuit simplification technique to generate several logically equivalent circuits for the Fredkin and Toffoli gates under all-to-all and linear qubit connectivity. In the former case, we started from the \textsc{cnot}-count-optimal circuits shown in Fig.~\ref{fig:standard-decomps}, so the obtained circuits had the same number of \textsc{cnot} gates, though arranged in a different way. Under linear connectivity constraints, our starting point also corresponded to the circuits in Fig.~\ref{fig:standard-decomps} but with the two \textsc{cnot}s between the outermost qubits requiring a \textsc{swap} before and after the execution of the actual \textsc{cnot}. This na\"{i}f approach to handle the qubit connectivity restrictions is naturally far from optimal, and therefore our ZX-calculus-based technique yielded circuits with a significantly lower number of \textsc{cnot}s. 

At the end of the search procedure, further logically equivalent circuits with different \textsc{cnot} structures were generated from the circuits directly obtained from the original ZX-calculus-based procedure by exploiting the symmetries of the Toffoli and Fredkin gates, namely the invariance of the former under permutations of all three qubits (once it is converted into a controlled-controlled-\textsc{z} (\textsc{ccz}) gate by applying a pair of Hadamard gates on either side, as discussed in Appendix \ref{AppB}), the invariance of the latter under the permutation of the two target-qubits, and the invariance of both under inversion (since the Fredkin and Toffoli gates are self-inverses). Under linear qubit connectivity, some of these transformations were discarded, as they resulted in \textsc{cnot} gates between unconnected qubits. All in all, this process allowed to increase the number of circuits with different entangling gate structures for each gate implementation.

The \textsc{cnot} count and the number of equivalent circuits for each scenario of qubit connectivity and position of the odd qubit (target-qubit for Toffoli and control-qubit for Fredkin) are shown in Table \ref{tab:stats}. When all qubits are connected to one another, the placement of the odd qubit is immaterial. For the Toffoli gate, even under linear connectivity, the position of the target-qubit is irrelevant as far as the entangling gate structure of the circuit is concerned, since the target-qubit can be changed by simply moving a pair of Hadamard gates, one on either end of the circuit. This follows from the close relation of the Toffoli gate to the \textsc{ccz} gate, which is invariant under permutations of the three qubits (see Appendix \ref{AppB}). Fig. \ref{fig:lin-decomps} shows an example of a circuit with the lowest \textsc{cnot} count for each of the three scenarios of linear qubit connectivity.

\begin{table}[t]
\begingroup\setlength{\fboxsep}{0pt}
\colorbox{gray!5}{\parbox{1\columnwidth}{
\begin{tabularx}{\columnwidth}{ *{6}{m{0.12\columnwidth}>{}m{0.2\columnwidth}>{\centering\arraybackslash}m{0.15\columnwidth}>{\centering}m{0.12\columnwidth}>{\centering\arraybackslash}m{0.16\columnwidth}>{\centering\arraybackslash}m{0.16\columnwidth}} }
\hline
\hline
\noalign{\smallskip}
 Gate & Connectivity & Odd qubit & \textsc{cnot} & No. equiv. & No. equiv. \\ 
 & & placement & count & circ. & circ. with & \\
 & & & & & +1 \textsc{cnot} & \\
\hline
\hline
\noalign{\smallskip}
Toffoli & $\begin{array}{l} \textrm{All-to-all} \\ \textrm{Linear}\end{array}$ & $\begin{array}{l} \textrm{Anywhere} \end{array}$ & $\begin{array}{r} \phantom{0}6 \\ \,8 \end{array}$ & $\begin{array}{r} 48 \\ 18 \end{array}$ & $\begin{array}{r} - \\ 54 \end{array}$ \\
\hline
\noalign{\smallskip}
Fredkin & $\begin{array}{l} \textrm{All-to-all} \\ \textrm{Linear} \\ \, \end{array}$ & $\begin{array}{l} \textrm{Anywhere} \\ \textrm{Ends} \\ \textrm{Center} \end{array}$ & $\begin{array}{r} 7 \\ 8 \\ 10 \end{array}$ & $\begin{array}{r} 40 \\ 8 \\ 2 \end{array}$ & $\begin{array}{r} - \\ 22 \\ 69 \end{array}$ \\
\hline
\hline
\end{tabularx}
}}\endgroup
\caption{\textsc{cnot} count and number of equivalent circuits generated for Fredkin and Toffoli decompositions in five different scenarios of qubit connectivity and position of odd qubit (control-qubit for the former and target-qubit for the latter). All circuits have been made available online in \textsc{qasm} file format. In addition to the optimal-\textsc{cnot}-count circuits, we also provide all circuits with linear qubit connectivity that have one more \textsc{cnot} gate than the optimal, as these may be useful for equivalent circuit averaging (see Section \ref{sec:ECA}).}
\label{tab:stats}
\end{table}

\begin{table}[t]
\begingroup\setlength{\fboxsep}{0pt}
\colorbox{gray!5}{\parbox{1\columnwidth}{
\begin{tabular}{p{0.32\columnwidth} c c c c c}
\hline
\hline
  & Toffoli & Toffoli & Fredkin & Fredkin & Fredkin \\ 
 & all-to-all & linear & all-to-all & linear & linear  \\
 & & & & (ends) & (center)  \\
\hline
\hline
\noalign{\smallskip}
Here & 6 & 8 & 7 & 8 & 10 \\ 
\texttt{BQSKit}~\cite{BQSKit2021} & 6 & 8 & 7 & 8 & 10 \\
\texttt{CPFlow}\cite{CPFlow2023} & 6 & 8 & 7 & 8 & 11 \\
\texttt{Qiskit}\cite{QiskitTranspile} & 6 & 10 & 7 & 11 & 17 \\
Duckering et al.~\cite{duckering2021orchestrated} & 6 & 8 & - & - & - \\
Liu et al.~\cite{liu2023qcontext} & 6 & 8 & - & - & - \\
\hline
\hline
\end{tabular}
}}\endgroup
\caption{\textsc{cnot} count of decompositions of Fredkin and Toffoli gates for five different scenarios of qubit connectivity and position of odd qubit, as in Table \ref{tab:stats}. The \texttt{BQSKit}\cite{BQSKit2021}, \texttt{CPFlow}\cite{CPFlow2023} and \texttt{Qiskit}\cite{QiskitTranspile} unitary decomposition methods were used to benchmark our results. The lowest \textsc{cnot} counts reported in the literature~\cite{duckering2021orchestrated, liu2023qcontext} for the Toffoli gate are also included for reference; no analogous results for the Fredkin gate could be found. Apart from achieving the lowest \textsc{cnot} count in all five cases, the multiple equivalent circuits we have generated have the additional benefits of being exact --- as all single-qubit-gate parameters are exact fractions of $\pi$ --- and having been stored in memory --- so that they can be retrieved when necessary, thus avoiding carrying out the unitary decomposition from scratch.}
\label{tab:benchmark}
\end{table}

The shallowest circuits for the Fredkin and Toffoli gates generated by our ZX-calculus-based unitary decomposition technique have the lowest \textsc{cnot} counts in the literature. Table \ref{tab:benchmark} shows the \textsc{cnot} counts achieved by different basis gate decomposition methods for the five different scenarios of qubit connectivity and odd qubit placement previously considered in Table \ref{tab:stats}. In addition, Table \ref{tab:benchmark} includes the \textsc{cnot} counts presented in two earlier papers~\cite{duckering2021orchestrated, liu2023qcontext} for the decomposition of the Toffoli gate under all-to-all and linear qubit connectivity; analogous results for the Fredkin gate could not be found in the literature. The lowest \textsc{cnot} counts herein reported have also been attained by the \texttt{BQSKit}~\cite{BQSKit2021} and \texttt{CPFlow}~\cite{CPFlow2023} packages (with the exception of the Fredkin gate under linear qubit connectivity and the control-qubit at the center in the latter case). Our results offer three advantages relative to using these alternative packages. First, the circuits we have generated have been decomposed in the $\{\textsc{cnot}, R_z(\theta), R_x(\theta)\}$ basis~\cite{nielsen2002quantum} with all single-qubit-gate parameters $\theta$ corresponding to exact fractions of $\pi$. Besides guaranteeing the decompositions are accurate to numerical precision, these circuits may also be useful for fault-tolerant quantum hardware, as the decomposition of single-qubit gates with respect to a finite basis is simplified. Second, instead of just one decomposition, our method generates several logically equivalent ones. Third, all equivalent circuits we have generated for the Fredkin and Toffoli gates have been made available online, so they can just be saved in memory and retrieved when required \cite{cruz2023Fredkin}.

Having different circuits that realize the same gate offers the possibility of implementing a number of methods that address the limitations of near-term quantum hardware. For example, two decompositions of the same gate may allow for a different degree of simplification of the circuit of which the gate is part by taking the context around the gate into account\cite{liu2023qcontext}. Likewise, if the \textsc{cnot} gate implemented between a pair of qubits has an especially high error rate, one may choose a circuit that makes use of the fewest number of \textsc{cnot}s between those two qubits to maximize the fidelity of the outcome. Even more importantly, it is possible to mitigate the effects of coherent errors through equivalent circuit averaging\cite{hashim2022optimized, campbell2017shorter, hastings2016turning}. Before we discuss this application in Section \ref{sec:ECA}, we will consider the implementation of the Fredkin and Toffoli gates when the three qubits are not adjacent.

\section{Decompositions for non-adjacent qubits \label{sec:nonadj-dec}}

In this section, we address the implementation of the Fredkin and Toffoli gates when the three active qubits are not adjacently connected. In this scenario, the neighboring qubits in their path must be used to implement
the global long-range unitary. Avoiding a direct basis gate decomposition, we introduce the \emph{cnot-swapping} method and show how it allows for an efficient rerouting of the qubits before and after applying the three-qubit circuits in Fig.~\ref{fig:lin-decomps}. We first examine the general applicability of this technique to moving any qubit with respect to which the matrix representation of the gate is diagonal in the computational basis. This includes the important case of control-qubits. Then, we explain how it can also be used to move the target-qubits in multi-controlled-\textsc{not} operations. The cases of a long-range \textsc{cnot} and the Toffoli gates follow immediately from these two instances. Lastly, the application to the Fredkin gate is discussed.

\subsection{CNOT-SWAP rerouting}

To introduce the \textsc{cnot-swap} gate, let us start by considering its action on a pair of classical bits,
\begin{center}
\begin{quantikz}[row sep={0.6cm,between origins}, column sep=0.3cm]
& \swap{1} & \qw \midstick[2,brackets=none]{$=$} & & \lstick{$\ket{i_1}$} & \ctrl{1} & \targ{} & \qw \rstick{$\ket{i_2}$} \\
& \targ{} & \qw & & \lstick{$\ket{i_2}$} & \targ{} & \ctrl{-1} & \qw \rstick{$\ket{i_1 \oplus i_2}$,}
\end{quantikz}
\end{center}
where $i_1,i_2 \in \left\{0,1\right\} $. In words, the \textsc{cnot-swap} uses the first bit to control a \textsc{not} operation on the second one, while perfectly moving the second bit into the state of the first. From the point of view of the second bit, the effective action of this gate is a \textsc{cnot}, whereas from the perspective of the first bit its effective action is a \textsc{swap},
hence the shorthand diagram. In the end, one of the states is left intact, or \emph{clean}, while the other accumulates computation, becoming \emph{dirty}. Operating on an arbitrary two-qubit state $\ket{\psi}=\sum_{i_1, i_2} a_{i_1 i_2}\ket{i_{1}}\otimes\ket{i_{2}}$ leads to
\begin{equation}
    \begin{aligned}
    \ket{\psi'} & =  \textsc{cnot-swap}_{1,2} \ket{\psi} \\
                & = \sum_{i_1,i_2}a_{i_1 i_2}\ket{i_{2}}\otimes\ket{i_{1} \oplus i_{2}}.
\end{aligned}
\label{eq:cnot-swap}
\end{equation}
As a result, and most importantly, only the computational basis elements are permuted, leaving the amplitudes unchanged.

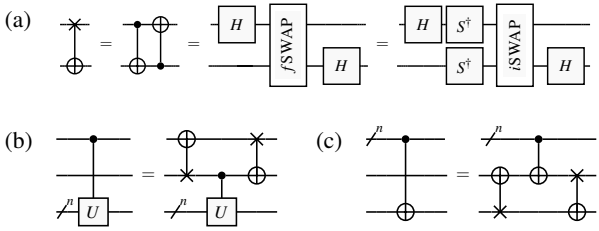
\begin{figure}[t]
\centering

\begin{tabular}{lll}
\multicolumn{3}{l}{(a) \tikzset{ operator/.append style={draw, fill=white, text height=45pt, text width=20pt}}
\raisebox{-0.76\totalheight}{
\begin{tikzpicture}
\node[scale=0.69]{
\begin{tikzcd}[row sep={0.8cm,between origins}, column sep=0.04cm]
& \qw & \qw & \swap{1} & \qw & \qw & \qw \midstick[2,brackets=none]{$=$} & \qw & \qw & \ctrl{1} & \qw & \qw & \targ{} & \qw & \qw & \qw & \qw \midstick[2,brackets=none]{$=$} & \qw & \qw & \qw & \qw & \qw & \gate[style={text height = 20pt, fill=gray!5}]{H} & \gate[label style={black, rotate=90, fill=gray!5, fill=gray!5}, wires=2]{\small{f\textsc{SWAP}}} & \arrow[arrows]{ll} & \qw & \qw & \qw & \qw & \qw & \qw \midstick[2,brackets=none]{$=$} & \qw & \qw & \qw & \qw & \gate[style={text height = 20pt, fill=gray!5}]{H} & \qw & \qw & \qw & \qw & \qw & \gate[style={text height = 20pt, fill=gray!5}]{S^\dagger} & \gate[label style={black, rotate=90, fill=gray!5}, wires=2]{\small{i\textsc{SWAP}}} & \arrow[arrows]{ll} & \qw & \qw & \qw & \qw & \qw & \qw \\
& \qw & \qw & \targ{} & \qw & \qw & \qw & \qw & \qw & \targ{} & \qw & \qw & \ctrl{-1} & \qw & \qw & \qw & \qw & \qw & \qw & \qw & \qw & \qw & \qw & \qw & \arrow[arrows]{ll} & \gate[style={text height = 20pt, fill=gray!5}]{H} & \qw & \qw & \qw & \qw & \qw & \qw & \qw & \qw & \qw & \qw & \qw & \qw & \qw & \qw & \qw & \gate[style={text height = 20pt, fill=gray!5}]{S^\dagger} & \qw & \arrow[arrows]{ll} & \gate[style={text height = 20pt, fill=gray!5}]{H} & \qw & \qw & \qw & \qw & \qw \\
\end{tikzcd}
};
\end{tikzpicture}
}}\tabularnewline
(b) \raisebox{-0.75\totalheight}{
\begin{tikzpicture}
\node[scale=0.745]{
\begin{tikzcd}[row sep={0.65cm,between origins}, column sep=0.2cm]
\qw & \qw & \ctrl{2} & \qw & \qw \midstick[3,brackets=none]{$=$} & \targ{} & \qw & \swap{1} & \qw \\
\qw & \qw & \qw & \qw & \qw & \swap{-1} & \ctrl{1} & \targ{} & \qw \\
\qw & \qwbundle{n} & \gate{U} & \qw & \qw & \qwbundle{n} & \gate{U} & \qw & \qw
\end{tikzcd}
};
\end{tikzpicture}
} &  & (c) \raisebox{-0.75\totalheight}{
\begin{tikzpicture}
\node[scale=0.745]{
\begin{tikzcd}[row sep={0.65cm,between origins}, column sep=0.18cm]
\qw & \qwbundle{n} & \qw & \ctrl{2} & \qw & \qw & \qw \midstick[3,brackets=none]{$=$} & \qwbundle{n} & \qw & \ctrl{1} & \qw & \qw & \qw \\
\qw & \qw & \qw & \qw & \qw & \qw & \qw & \targ{} & \qw & \targ{} & \qw & \swap{1} & \qw \\
\qw & \qw & \qw & \targ{} & \qw & \qw & \qw & \swap{-1} & \qw & \qw & \qw & \targ{} & \qw
\end{tikzcd}
};
\end{tikzpicture}
}\tabularnewline
\end{tabular}
\caption{(a) Shorthand diagram for the \textsc{cnot-swap}, which is equivalent up to single-qubit transformations to the \emph{fermionic} \textsc{swap}\cite{verstraete2009quantum} and \emph{i}\textsc{swap}\cite{williams2011quantum} gates. The \textsc{cnot-swap} gate was also discussed previously under the name ``double-\textsc{cnot}'' and shown to be a maximally non-local operator\cite{collins2001nonlocal}. (b) One-hop movement of the control-qubit of an arbitrary controlled gate via two \textsc{cnot-swap}s. (c) One-hop movement of the target-qubit of a multi-controlled-\textsc{not} gate via two \textsc{cnot-swap}s. Note that the direction of the \textsc{cnot-swap}s is reversed with respect to the rerouting of a control-qubit shown in (b).}
\label{fig:cnot-swap}
\end{figure}

Let us now suppose that we wish to implement a two-qubit gate $V$ between two non-adjacent qubits. The general approach would be to bring the two qubits together through a network of \textsc{swap} gates, apply $V$ locally to a pair of adjacent qubits, and finally reverse the initial \textsc{swap} network to return the qubits to their original positions.
However, provided that $V$ is diagonal in the computational basis of the moving qubit, a more efficient alternative is possible by replacing every \textsc{swap} with a \textsc{cnot-swap}, thereby saving two \textsc{cnot}s for every qubit hop and its reversal (see Fig.~\ref{fig:cnot-swap}(b)). The moving qubit is the clean qubit of every \textsc{cnot-swap}. Although the qubits it goes past are initially left dirty, the final \textsc{cnot-swap} network cleans them to recover their original form.
This is possible because $V$ is guaranteed not to change the computational basis states of the moving qubit. Hence, after its application, each computational basis state of a dirty qubit is still associated with the computational basis state of the moving qubit responsible for its garbage (see Eq.~\ref{eq:cnot-swap}), and it can be cleaned by uncomputing the initial \textsc{cnot-swap} network. More generally, this method is valid to reroute any qubit on which a given $n$-qubit gate has support, provided that this unitary only modifies its amplitudes up to a relative phase factor.

\subsection{Long-range CNOT and Toffoli gates}

Let us now consider the important case of rerouting a control-qubit, as illustrated in Fig. \ref{fig:cnot-swap}(b). An $\ket{i_1}\otimes \ket{i_2}$ basis state of the top two qubits is first transformed into $\ket{i_1\oplus i_2} \otimes \ket{i_1}$ by \textsc{cnot-swap}$_{2,1}$. The subsequent controlled-operation on the bottom $n+1$ qubits therefore becomes controlled by $\ket{i_1}$ and preserves this state, as intended. Finally, by reversing the direction of the \textsc{cnot-swap}, the top two-qubit state is transformed back into $\ket{i_1} \otimes \ket{i_1 \oplus i_1 \oplus i_2} = \ket{i_1} \otimes \ket{i_2}$. Longer movements of the control are clearly generalized by the sequential application of this process.

The \textsc{cnot-swap} can also be used to move the target-qubit of a multi-controlled-\textsc{not} gate (\textsc{mcx}), as shown in Fig.~\ref{fig:cnot-swap}(c). The crucial difference relative to the previously considered case of a control-qubit is that the moving target-qubit is the dirty qubit of the \textsc{cnot-swap}, while the idle qubits it goes past are left clean. This is why the \textsc{cnot-swap} gates have opposite orientations with respect to the direction of flow of the moving qubit in Figs. \ref{fig:cnot-swap}(b)-(c). In the circuit of Fig.~\ref{fig:cnot-swap}(c), the basis state $\ket{i_{n+1}} \otimes \ket{i_{n+2}}$ of the bottom two qubits is transformed into \mbox{$\ket{i_{n+1}\oplus i_{n+2}} \otimes \ket{i_{n+1}}$} by the first \textsc{cnot-swap} gate; the \textsc{mcx} operation yields the state $\ket{(\bigoplus_{j=1}^{n}i_{j}) \oplus (i_{n+1}\oplus i_{n+2})} \otimes \ket{i_{n+2}}$; finally, the \textsc{cnot-swap}$_{n+1,n+2}$ gate transforms the state of the bottom two qubits into $\ket{i_{n+2}} \otimes \ket{(\bigoplus_{j=1}^{n}i_{j}) \oplus i_{n+1} \oplus (i_{n+2} \oplus i_{n+2}) } = \ket{i_{n+2}} \otimes \ket{ (\bigoplus_{j=1}^{n}i_{j}) \oplus i_{n+1} }$, as expected for a \textsc{mcx} gate.

\begin{figure}[t]
\centering
\begin{tabular}{lll}
(a)\;\;\raisebox{-0.85\totalheight}{
\begin{tikzpicture}[every node/.style={thick, circle, inner sep=1pt, scale=0.5}, node distance=0, on grid, auto, thick, scale=0.68, rotate=0]
     \node[state, fill=blue!15] at  (0, 3)   (1) 	{\Large 1};
     \node[state, fill=red!15] at  (1, 3)   (2) 	{\Large 2};
     \node[state]               at  (2, 3)   (3) 	{\Large 3};
     \node[state]               at  (0, 2)   (4) 	{\Large 4};
     \node[state, fill=red!15] at  (1, 2)   (5) 	{\Large 5};
     \node[state]               at  (2, 2)   (6) 	{\Large 6};
     \node[state]               at  (0, 1)   (7) 	{\Large 7};
     \node[state, fill=red!15] at  (1, 1)   (8) 	{\Large 8};
     \node[state, fill=blue!15] at  (2, 1)   (9) 	{\Large 9};
     \path[-]    (1)   edge   node   {}  (2)
                 (2)   edge   node   {}  (3)
                 (1)   edge   node   {}  (4)
                 (4)   edge   node   {}  (5)
                 (4)   edge   node   {}  (7)
                 (5)   edge   node   {}  (6)
                 (5)   edge   node   {}  (2)
                 (5)   edge   node   {}  (8)
                 (6)   edge   node   {}  (3)
                 (6)   edge   node   {}  (9)
                 (7)   edge   node   {}  (8)
                 (8)   edge   node   {}  (9);
\end{tikzpicture}
} &  & (b)\raisebox{-0.8\totalheight}{
\begin{tikzpicture}
\node[scale=0.75]{
\begin{tikzcd}[row sep={0.5cm,between origins}, column sep=0.11cm]
    \lstick{$\textcolor{blue}{1}$} & \ctrl{4} & \qw \midstick[5,brackets=none]{$=$} & \targ{} & \ctrl{1} & \qw & \qw & \qw & \qw & \qw & \qw & \ctrl{1} & \targ{} & \qw \\
    \lstick{$\textcolor{red}{2}$} & \qw & \qw \qw & \ctrl{-1} & \targ{} & \qw & \qw & \ctrl{1} & \qw & \ctrl{1} & \qw & \targ{} & \ctrl{-1} & \qw \\
    \lstick{$\textcolor{red}{5}$} & \qw & \qw \qw & \qw & \qw & \qw & \ctrl{1} & \targ{} & \ctrl{1} & \targ{} & \qw & \qw & \qw & \qw \\
    \lstick{$\textcolor{red}{8}$} & \qw & \qw \qw & \targ{} & \ctrl{1} & \qw & \targ{} & \qw & \targ{} & \qw & \qw & \ctrl{1} & \targ{} & \qw \\
    \lstick{$\textcolor{blue}{9}$} & \targ{} & \qw & \ctrl{-1} & \targ{} & \qw & \qw & \qw & \qw & \qw & \qw & \targ{} & \ctrl{-1} & \qw
\end{tikzcd}
};
\end{tikzpicture}
}\tabularnewline
\multicolumn{3}{l}{(c)\raisebox{-0.85\totalheight}{
\begin{tikzpicture}
\node[scale=0.71]{
\begin{tikzcd}[row sep={0.4cm,between origins}, column sep=0.11cm]
    & \ctrl{12} & \qw \midstick[13,brackets=none]{$=$} & \targ{} & \qw \gategroup[3,steps=2,style={dashed, rounded corners, draw=red, inner xsep=0pt, inner ysep=0pt}, background]{} & \ctrl{1} & \qw & \qw & \qw & \qw & \qw & \qw & \qw & \qw & \qw & \qw & \qw & \qw & \qw & \qw & \qw & \qw & \qw & \qw & \ctrl{1} & \qw & \targ{} & \qw \\
    & \qw & \qw \qw & \ctrl{-1} & \targ{} & \targ{} & \ctrl{1} & \qw & \qw & \qw & \qw & \qw & \qw & \qw & \qw & \qw & \qw & \qw & \qw \qw & \qw & \qw & \qw & \qw & \ctrl{1} & \targ{} & \targ{} & \ctrl{-1} & \qw \\
    & \qw & \qw \qw & \qw & \ctrl{-1} & \targ{} & \targ{} & \ctrl{1} & \qw & \qw & \qw & \qw & \qw & \qw & \qw & \qw & \qw & \qw & \qw & \qw & \qw & \qw & \ctrl{1} & \targ{} & \targ{} & \ctrl{-1} & \qw & \qw \\
    & \qw & \qw & \qw & \qw & \ctrl{-1} & \targ{} & \targ{} & \ctrl{} & \qw & \qw & \qw & \qw & \qw & \qw & \qw & \qw & \qw \qw & \qw & \qw & \qw & \ctrl{} & \targ{} & \targ{} & \ctrl{-1} & \qw & \qw & \qw & \\
    & & & & & & & & & & \lstick{$^{\ddots}\phantom{.......}^{\ddots}$} & & & & & & & &  & \rstick{$^{\iddots}\phantom{.......}^{\iddots}$} & \\
    & \qw & \qw \qw & \qw & \qw & \qw & \qw &  \ctrl{} & \qw & \qw & \targ{} & \qw &  \qw & \qw \gategroup[3,steps=4,style={dashed, rounded corners, draw=blue, inner xsep=0pt, inner ysep=0pt}, background]{} & \ctrl{1} & \qw & \ctrl{1} & \qw & \qw & \targ{} & \qw & \qw & \ctrl{} & \qw & \qw & \qw & \qw & \qw  \\
    & \qw & \qw \qw & \qw & \qw & \qw & \qw & \qw & \qw & \qw & \qw & \qw & \qw & \ctrl{1} & \targ{} & \ctrl{1} & \targ{} & \qw & \qw & \qw & \qw & \qw & \qw & \qw & \qw & \qw & \qw & \qw \\
    & \qw & \qw \qw & \qw & \qw & \qw & \qw & \targ{} & \qw & \qw & \ctrl{} & \qw & \qw & \targ{} & \qw & \targ{} & \qw & \qw & \qw & \ctrl{} & \qw & \qw & \targ{} & \qw & \qw & \qw & \qw & \qw \\
    & & & & & & & & & & \lstick{$^{\iddots}\phantom{.......}^{\iddots}$} & & & & & & & &  & \rstick{$^{\ddots}\phantom{.......}^{\ddots}$} & \\
    & \qw & \qw & \qw & \qw & \targ{} & \ctrl{} & \ctrl{1} & \targ{} & \qw & \qw & \qw & \qw & \qw & \qw & \qw & \qw & \qw & \qw & \qw & \qw & \targ{} & \ctrl{1} & \ctrl{} & \targ{} & \qw & \qw & \qw & \\
    & \qw & \qw & \qw & \targ{} & \ctrl{-1} & \ctrl{1} & \targ{} & \qw & \qw & \qw & \qw & \qw & \qw & \qw & \qw & \qw & \qw & \qw & \qw & \qw & \qw & \targ{} & \ctrl{1} & \ctrl{-1} & \targ{} & \qw & \qw  \\
    & \qw & \qw \qw & \targ{} & \ctrl{-1} & \ctrl{1} & \targ{} & \qw & \qw & \qw & \qw & \qw & \qw & \qw & \qw & \qw & \qw & \qw & \qw & \qw & \qw & \qw & \qw & \targ{} & \ctrl{1} & \ctrl{-1} & \targ{} & \qw \\
    & \targ{} & \qw & \ctrl{-1} & \qw & \targ{} & \qw & \qw & \qw & \qw & \qw & \qw & \qw & \qw & \qw & \qw & \qw & \qw & \qw & \qw & \qw & \qw & \qw & \qw & \targ{} & \qw & \ctrl{-1} & \qw 
\end{tikzcd}
};
\end{tikzpicture}
}}\tabularnewline
\end{tabular}
\caption{Long-range \textsc{cnot} gate circuit. (a) A $3\times3$ square qubit lattice layout example with nearest-neighbor connections only. The five qubits on which the long-range \textsc{cnot} operates are highlighted in color. Active qubits, in blue, are connected through idle qubits, in red, across one of the shortest paths in terms of the Manhattan distance. (b) Decomposition of the long-range \textsc{cnot} gate between qubits 1 and 9, which are both moved towards each other to minimize circuit depth. (c) General construction of a long-range \textsc{cnot} gate, minimizing \textsc{cnot} count and depth, in this order. The subcircuit inside the blue box corresponds to the optimal decomposition of a \textsc{cnot} with an idle qubit between control and target. For a greater number of idle qubits between the pair of active qubits, networks of \textsc{cnot-swap}s on either side are applied. The permutation of \textsc{cnot}s from adjacent \textsc{cnot-swap}s, highlighted in the red box for the second \textsc{cnot} of the first qubit pair and the first \textsc{cnot} in the second qubit pair, allows each rerouting layer to fit two to four gates.}
\label{fig:long-cnot}
\end{figure}
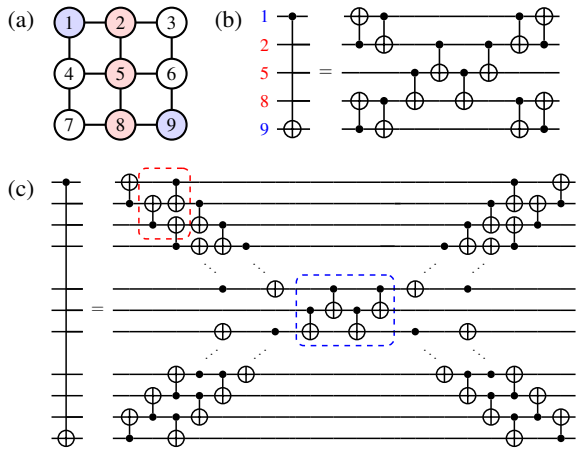

In constructing \textsc{cnot-swap} networks to facilitate extended movements of control and target qubits, a simple but important simplification can be applied to the resultant circuits. Specifically, the last \textsc{cnot} in a given \textsc{cnot-swap} can be permuted with the initial \textsc{cnot} of the subsequent \textsc{cnot-swap} along each network path. This interchange is feasible as these \textsc{cnot} gates lack a common qubit serving as the target for one and the control for the other. This rearrangement allows some pairs of \textsc{cnot} gates along each of the network paths to be applied concurrently, thereby achieving further reduction in circuit depth. For a visual representation, refer to Fig.~\ref{fig:long-cnot}(c).

The minimal case of a single control-qubit results in the so-called \emph{long-range} \textsc{cnot}, i.e., a \textsc{cnot} gate acting on two qubits that are not directly connected to each other. Applying the \textsc{cnot-swap} methodology herein introduced to the long-range \textsc{cnot} gate decomposition produces both the lowest number of \textsc{cnot} gates and circuit depth, in this sequential order, in the literature.

A brief review of the literature on the implementation of the long-range \textsc{cnot} is in order. The standard approach to the synthesis of a long-range \textsc{cnot} gate from basic circuit primitives amounts to the introduction of \textsc{swap} gates along the shortest path connecting the control and target qubits, resulting in their adjacent placement, at which point a \textsc{cnot} gate can be directly applied. With $n \geq 1$ intermediary qubits between the control-qubit and target-qubit, this method results in a circuit comprising $6n+1$ \textsc{cnot} gates, with a best-case depth of $\sim 3n$, assuming that the control-qubit and target-qubit of the long-range \textsc{cnot} are both moved towards each other in parallel. An improvement over this simple \textsc{swap}-based method was proposed by Shende \textit{et al.} \cite{shende2006synthesis}; the number of \textsc{cnot} gates was reduced to $4n$ at the expense of increasing circuit depth to $4n$ as well. Interestingly, this method appears to have been re-discovered recently with an algorithm based on the cryptographic problem of syndrome decoding\cite{brugiere2020quantum}. Later, Kutin \emph{et al.}\cite{kutin2007computation} proposed a circuit construction  reducing circuit depth to $\sim n$ while increasing \textsc{cnot} count in only 1 unit relative to the circuit by Shende \emph{et al.}.

The decomposition of the long-range \textsc{cnot} that we arrive to using the \textsc{cnot-swap} methodology and represent in Fig.~\ref{fig:long-cnot} reaches the circuit depth\footnote{To be precise, the decomposition represented in Fig.~\ref{fig:long-cnot} has depth $n+7$ for $n\geq4$, only 1 unit away from the $n+6$ depth that \cite{kutin2007computation} can guarantee for arbitrary $n$ (cf. theorem 2.1).} of $\sim n$ from Kutin \emph{et al.} while maintaining exactly the same minimal number of $4n$ \textsc{cnot}s achieved by Shende \emph{et al}. We have verified its optimality for the cases where the two active qubits are separated by $n=1$ and $n=2$ idle qubits, minimizing these instances primarily by \textsc{cnot} count and secondarily by depth; in both instances an exhaustive search was carried out with a gate set containing only nearest-neighbor \textsc{cnot} gates.

Furthermore, the parallelized structure of the circuit provides additional advantages, since composing two long-range \textsc{cnot}s one after the other, possibly interposed by some local operations, allows for a further simplification of the overall circuit by canceling out subsequent \textsc{cnot}s on the same qubit pairs. An important case where this occurs is in sequences of \textsc{cnot}s with a fixed control-qubit but multiple target-qubits, which are commonly found in state distillation and error correction \cite{fowler2012surface, devitt2013quantum}. Another relevant instance of the use of \textsc{cnot-swap}s to reduce the depth and \textsc{cnot} count of quantum circuits is the implementation of complex exponentials of Pauli strings, which are ubiquitous in Hamiltonian simulation \cite{Tacchino2020}. An example for each of these cases is given in Appendix \ref{AppC}.

To synthesise the Toffoli gate on a trio of non-adjacent qubits, interpreting each \textsc{cnot} that appears in the decomposition as a long-range \textsc{cnot} may not be the most advantageous solution. However, and most importantly, the same underlying ideas discussed before can be applied to bring the qubits together and implement the Toffoli gate through the circuits introduced in Section \ref{sec:adj-dec} that assume linear qubit connectivity. Concretely, both control-qubits and the target-qubit of the Toffoli gate can be moved similarly to the control-qubit and target-qubit of the long-range \textsc{cnot}, respectively. For the sake of clarity, \mbox{Fig. \ref{fig:toffoli_fredkin_non_adj}(a)} illustrates a specific example of this \textsc{cnot-swap}-based decomposition for a Toffoli gate. As far as we are aware, this decomposition has not appeared in the literature before.

For the long-range \textsc{cnot} and Toffoli gates, all \textsc{swap}s can be replaced with \textsc{cnot-swap}s in the qubit rerouting layers before and after the actual gate. Hence, if the cumulative number of idle qubits that are gone past by the three (two) qubits on which the Toffoli (long-range \textsc{cnot}) gate acts nontrivially is $n$, the \textsc{cnot} count of the rerouting networks is reduced from $6n$ to $4n$, and their depth is reduced from $\sim6n$ to $\sim n$.

\subsection{Fredkin gate}

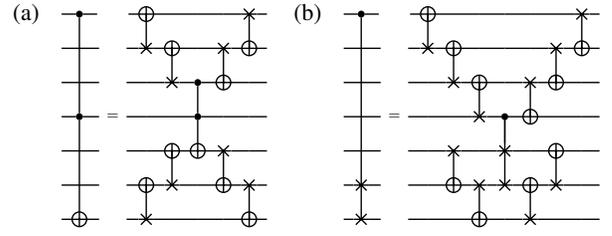
\begin{figure}[t]
\centering
\begin{tabular}{ll}
(a) \raisebox{-0.9\totalheight}{
\begin{tikzpicture}
\node[scale=0.65]{
\begin{tikzcd}[row sep={0.7cm,between origins}, column sep=0.2cm]
& \ctrl{3} & \qw \midstick[7,brackets=none]{$=$} & \targ{} & \qw & \qw & \qw & \swap{1} & \qw \\
& \qw & \qw & \swap{-1} & \targ{} & \qw & \swap{1} & \targ{} & \qw  \\
& \qw & \qw & \qw & \swap{-1} & \ctrl{1} & \targ{} & \qw & \qw  \\
& \ctrl{3} & \qw & \qw & \qw & \ctrl{1} & \qw & \qw & \qw  \\
& \qw & \qw & \qw & \targ{} & \targ{} & \swap{1} & \qw & \qw  \\
& \qw & \qw & \targ{} & \swap{-1} & \qw & \targ{} & \swap{1} & \qw  \\
& \targ{} & \qw & \swap{-1} & \qw & \qw & \qw & \targ{} & \qw 
\end{tikzcd}
};
\end{tikzpicture}
} & (b) \raisebox{-0.9\totalheight}{
\begin{tikzpicture}
\node[scale=0.65]{
\begin{tikzcd}[row sep={0.7cm,between origins}, column sep=0.2cm]
& \ctrl{4} & \qw \midstick[7,brackets=none]{$=$} & \targ{} & \qw & \qw & \qw & \qw & \qw & \swap{1} & \qw \\
& \qw & \qw & \swap{-1} & \targ{} & \qw & \qw & \qw & \swap{1} & \targ{} & \qw  \\
& \qw & \qw & \qw & \swap{-1} & \targ{} & \qw & \swap{1} & \targ{1} & \qw & \qw  \\
& \qw & \qw & \qw & \qw & \swap{-1} & \ctrl{1} & \targ{} & \qw & \qw & \qw  \\
& \qw & \qw & \qw & \swap{1} & \qw & \swap{-1} & \qw & \targ{} & \qw & \qw  \\
& \swap{-1} & \qw & \qw & \targ{} & \swap{1} & \swap{-1} & \targ{} & \swap{-1} & \qw & \qw  \\
& \swap{-1} & \qw & \qw & \qw & \targ{} & \qw & \swap{-1} & \qw & \qw & \qw
\end{tikzcd}
};
\end{tikzpicture}
}
\end{tabular}
\caption{Shallow implementation of Toffoli (a) and Fredkin (b) gates when the three qubits on which they act are not adjacent in an architecture with linear connectivity constraints. To reroute the qubits, every \textsc{swap} gate was replaced by a \textsc{cnot-swap}, saving one \textsc{cnot} in each instance. This use of \textsc{cnot-swap}s to reroute the target-qubits of the Fredkin gate only works when both are moved past the same idle qubits, as discussed in the main text. This strategy of moving both the control-qubit and the pair of target-qubits of the Fredkin in parallel aims to minimize the circuit depth; we could instead move only the control through \textsc{cnot-swap} networks, which would achieve a lower overall \textsc{cnot} count, though at the cost of a greater depth. To fully appreciate the depth savings, consider the \textsc{cnot-swap} decomposition in terms of its constituent \textsc{cnot}s and the permutation trick depicted in Fig.~\ref{fig:long-cnot}(c).}
\label{fig:toffoli_fredkin_non_adj}
\end{figure}

Regarding the Fredkin gate, the control-qubit can always be moved through \textsc{cnot-swap} networks in a similar way to the control-qubits of the long-range \textsc{cnot} and Toffoli gates. As for the target-qubits, at first glance it appears that rerouting via \textsc{cnot-swap} networks is not a valid option, as the effective action of the Fredkin gate on the target-qubits is neither diagonal in the computational basis nor equivalent to a \textsc{not} gate. In any case, it is possible to apply the network of \textsc{cnot-swap}s (just like for the control-qubits of the long-range \textsc{cnot} and Toffoli gates) to the target-qubits of the Fredkin gate if they are moved together, as illustrated with an example in Fig. \ref{fig:toffoli_fredkin_non_adj}(b).

If the two target-qubits of the Fredkin gate are initially adjacent, they have to move past exactly the same idle qubits to reach the control-qubit, so the garbage introduced in the idle qubits can still be cleaned even if the two target-qubits are swapped by the Fredkin gate between the two rerouting layers. Taking the example shown in Fig. \ref{fig:toffoli_fredkin_non_adj}(b), let us consider the action of the networks of \textsc{cnot-swap}s on the computational basis states of all qubits before and after the Fredkin:
\begin{figure}[h]
\centering
\begin{tabular}{lll}
\raisebox{-0.9\totalheight}{
\begin{tikzpicture}
\node[scale=0.77]{
\begin{tikzcd}[row sep={0.7cm,between origins}, column sep=0.2cm]
\lstick{$\ket{1}$}   & \targ{} & \qw & \qw & \qw \\
\lstick{$\ket{i_1}$} & \swap{-1} & \targ{} & \qw & \qw  \\
\lstick{$\ket{i_2}$} & \qw & \swap{-1} & \targ{} & \qw  \\
\lstick{$\ket{i_3}$} & \qw & \qw & \swap{-1} & \qw  \\
\lstick{$\ket{i_4}$} & \qw & \swap{1} & \qw & \qw  \\
\lstick{$\ket{t_1}$} & \qw & \targ{} & \swap{1} & \qw  \\
\lstick{$\ket{t_2}$} & \qw & \qw & \targ{} & \qw
\end{tikzcd}
};
\end{tikzpicture}
} & \raisebox{-0.9\totalheight}{
\begin{tikzpicture}
\node[scale=0.77]{
\begin{tikzcd}[row sep={0.7cm,between origins}, column sep=0.2cm]
\lstick{$\ket{i_1 \oplus 1}$} & \qw & \qw \\
\lstick{$\ket{i_2 \oplus 1}$} & \qw & \qw  \\
\lstick{$\ket{i_3 \oplus 1}$} & \qw & \qw  \\
\lstick{$\ket{1}$}            & \ctrl{1} & \qw  \\
\lstick{$\ket{t_1}$}          & \swap{-1} & \qw  \\
\lstick{$\ket{t_2}$}          & \swap{-1} & \qw  \\
\lstick{$\ket{i_4 \oplus t_1 \oplus t_2}$} & \qw & \qw
\end{tikzcd}
};
\end{tikzpicture}
} & \raisebox{-0.9\totalheight}{
\begin{tikzpicture}
\node[scale=0.77]{
\begin{tikzcd}[row sep={0.7cm,between origins}, column sep=0.2cm]
\lstick{$\ket{i_1 \oplus 1}$} & \qw & \qw & \swap{1} & \qw \rstick{$\ket{1}$} \\
\lstick{$\ket{i_2 \oplus 1}$} & \qw & \swap{1} & \targ{} & \qw \rstick{$\ket{i_1}$} \\
\lstick{$\ket{i_3 \oplus 1}$} & \swap{1} & \targ{} & \qw & \qw \rstick{$\ket{i_2}$} \\
\lstick{$\ket{1}$}            & \targ{} & \qw & \qw & \qw \rstick{$\ket{i_3}$} \\
\lstick{$\ket{t_2}$}          & \qw & \targ{} & \qw & \qw \rstick{$\ket{i_4}$} \\
\lstick{$\ket{t_1}$}          & \targ{} & \swap{-1} & \qw & \qw \rstick{$\ket{t_2}$} \\
\lstick{$\ket{i_4 \oplus t_1 \oplus t_2}$} & \swap{-1} & \qw & \qw & \qw \rstick{$\ket{t_1}$}
\end{tikzcd}
};
\end{tikzpicture}
}
\end{tabular}
\end{figure}
\noindent Since the case where the control-qubit of the Fredkin gate is in state $\ket{c} = \ket{0}$ is trivial, we shall assume that the control-qubit is in state $\ket{c} = \ket{1}$, in which case the target-qubits $\ket{t_1}$ and $\ket{t_2}$ are swapped. The basis states of the idle qubits are represented as $\{ \ket{i_n} \}_{n=1}^{4}$. After the Fredkin gate swaps $\ket{t_1}$ and $\ket{t_2}$, both target-qubits are moved past the same idle qubits, so the undesired change in the latter that the former left jointly in the first network of \textsc{cnot-swap}s will still be reversed by the second network. Conversely, if the two target-qubits of the Fredkin gate are not next to each other, the \textsc{cnot-swap} gate cannot replace the \textsc{swap} gate in general. However, even if the two target-qubits are originally separated from each other, we may consider moving one of them (namely the one that is farthest from the control-qubit) towards the other via a network of \textsc{swap}s, and then move the pair of target-qubits together towards the control-qubit via a network of \textsc{cnot-swap}s. Meanwhile, the control qubit should also be moved towards the target qubits via a network of \textsc{cnot-swap}s to parallelize the rerouting, thus reducing the circuit depth.

Rerouting only the control qubit stands out as the best approach for minimizing the \textsc{cnot} count of the Fredkin gate when only the control qubit is non-adjacent to the targets. While targeting circuit depth reduction, however, moving only the control qubit yields a depth scaling of $\sim 2n$ in the rerouting networks, where $n$ is the number of idle qubits, whereas moving all three qubits concurrently into an intermediate position reaches $\sim n$ depth\footnote{Concretely, the depth of the rerouting network that moves only the control qubit is $2n+4$ for $n\geq 2$ idle qubits between the control and both targets, while the depth achieved by the network that moves the three qubits simulaneously, considering the targets begin in an adjacent position, is given by $n+\frac{15+(-1)^{n}}{2}$, for $n\geq 6$, when the pattern is fully developed.}. In the latter case, the strategy consists in hopping the control qubit by 1 position if $n=1$, or by $n+1-\left\lfloor \frac{n}{2} \right\rfloor$ positions if $n>1$ while also moving the target qubits together in the opposite direction to make them adjacent to the control. As a result, rerouting all three qubits at the same time may be the best option for minimizing environmental interactions in near-term quantum hardware or reducing total execution time in fault-tolerant hardware.

\section{Equivalent circuit averaging \label{sec:ECA}}

Exploiting the full potential of quantum computing and achieving super-polynomial algorithmic speedups will require further technological advancements that allow for the faithful execution of arbitrarily long quantum circuits. On both near- and long-term quantum hardware, this is hampered by two primary challenges: decoherence, which limits the amount of time during which quantum circuits can operate before incoherent errors accumulate, and control errors, which often arise from coherent sources. It still remains unclear which of these limitations will be harder to overcome. This is because there is typically a trade-off: deepening circuits enhances decoherence, while introducing parallelized operations to reduce depth simultaneously adds coherent noise.

Coherent noise sources may be more damaging to the operation of a digital quantum computer as their worst-case error rate scales as the square root of the average error rate, thus potentially leading to a faster deterioration of the fidelity of the outcome of a quantum circuit \cite{Iverson2020, Cai2020}. As a result, it is imperative to suppress coherent errors in gate implementations as much as possible and prevent their accumulation during the algorithmic execution, as it can incur constructive or destructive interference and lead to computational results that, while precise, are incorrect. In fact, coherent errors can be statistically resolved in the outcomes of current superconducting quantum processors even with very shallow circuits \cite{cruz2021testing}.

To this end, various methods that add new or modify existing single-qubit gates in the default circuit have been introduced. Important examples include dynamical decoupling \cite{tripathi2022suppression}, arbitrarily accurate composite pulse sequences \cite{brown2004arbitrarily}, and randomization procedures such as Pauli twirling \cite{geller2013efficient}, Pauli frame randomization \cite{ware2021experimental}, and randomized compiling \cite{wallman2016noise}. Another strategy consists in synthesizing close but different unitaries in such a way that mixing and averaging over them produces statistics closer to that of the target unitary \cite{hastings2016turning, campbell2017shorter}. The equivalent circuit averaging (ECA) technique \cite{hashim2022optimized} follows a similar spirit: different but logically equivalent circuits are executed, and their measurement statistics are aggregated to convert the different systematic errors into stochastic noise.

In this section, we test how the diversity of optimized circuits introduced in Section~\ref{sec:adj-dec} for the Fredkin and Toffoli gates allows to mitigate coherent errors via an ECA methodology. Concretely, the protocol we propose for the execution of quantum algorithms with Fredkin or Toffoli gates consists in building $M$ different but logically equivalent circuits and combining the measurement statistics from all of them. The available $S$ shots are evenly distributed through the $M$ equivalent circuits by measuring each of these $s=S/M$ times. To construct each of these circuits, a different unitary decomposition of the gates is applied each time the said gate appears in the circuit by uniformly sampling from our set of logically equivalent decompositions (the counts of which are summarized in Table~\ref{tab:stats}). In the presence of systematic control errors in the native gates of the quantum processor, each logically equivalent circuit will result in a slightly different unitary $C_i$ from the target unitary $T$. The resulting protocol can then be modeled as a uniform combination of $M$ unitary channels, configuring a \emph{uniformly-mixed-unitary} channel that transforms an input state $\rho$ according to
\begin{equation}
{\mathcal E}\left(\rho\right)=\frac{1}{M}\sum_{i=1}^{M}C_{i}\rho C_{i}^{\dagger}.
\label{eq:eca_channel}
\end{equation} 

In contrast to previously proposed ECA protocols, we recognize that the primary source of control errors in current quantum processors originates from the implementation of two-qubit gates rather than single-qubit gates. Consequently, our focus is directed towards devising a set of equivalent circuits featuring a variety of entangling gate structures. The greater the diversity of equivalent circuits, the more effective the ECA methodology is at mitigating the coherent errors.

\subsection{Approximating ideal and faulty circuits}

While the systematic nature of coherent errors makes it theoretically possible to correct them through recalibration or compensation operations, in practice, characterizing these errors on multi-qubit processors is an unmanageable task. The challenge stems from the lack of efficient methods to fully characterize the coherent processes that occur in all qubits in a timely manner when a single- or two-qubit gate is applied.

Similarly, without knowledge of the error processes in the device, it is not possible to know in advance which of the $M$ equivalent circuits is least impacted by them. Therefore, we propose the ECA procedure to produce a channel ${\mathcal E}$ (see \mbox{Eq. (\ref{eq:eca_channel})}) that achieves a better approximation, on average, to the target unitary than any individual $C_i$, as quantified by
\begin{equation}
d_{\diamondsuit}(\mathcal{E},\mathcal{T})\leq\left\langle d_{\diamondsuit}\left(\mathcal{C}_{i},\mathcal{T}\right)\right\rangle =\frac{1}{M}\sum_{i=1}^{M}d_{\diamondsuit}\left(\mathcal{C}_{i},\mathcal{T}\right),
\label{eq:eca_hypothesis}
\end{equation}
where $\mathcal{T}\left(\rho\right)=T\rho T^{\dagger}$ and $\mathcal{C}_i\left(\rho\right)=C_i\rho C_i^{\dagger}$ represent the quantum channels associated with the target unitary $T$ and the unitary corresponding to an equivalent circuit $C_i$, respectively, and $d_{\diamondsuit}$ is the diamond distance between two completely positive trace-preserving maps $\mathcal{M}$ and $\mathcal{M}'$, given by
\begin{align}
d_{\diamondsuit}(\mathcal{M},\mathcal{M}') & :=\|\mathcal{M}-\mathcal{M}'\|_{\diamondsuit} \nonumber \\
 & =\sup_{\rho}\left\Vert \left(\mathcal{M}\otimes \mathcal{I}\right)\rho-\left(\mathcal{M}'\otimes \mathcal{I}\right)\rho\right\Vert _{1}.
 \label{eq:diamond-dist}
\end{align}
Here, $\left\Vert \cdot\right\Vert _{1}$ is the trace norm and $\mathcal{I}$ is the identity map of the same dimensionality $n$ as $\mathcal{M}$ and $\mathcal{M}'$. The supremum is taken over all $n^2$-dimensional density matrices $\rho$. Geometrically, $0\leq d_{\diamondsuit}\leq2$ measures the maximum distinguishability (evaluated in terms of the trace distance) between the output states of the two maps under any input state. In other words, it quantifies the worst-case difference between the output states of the maps for any input state.

In order to assess how well Eq.~(\ref{eq:eca_hypothesis}) might hold in practice for the set of equivalent circuits we generated for the Fredkin and Toffoli gates, a concrete coherent-error model must be considered. We adopted a model recently introduced by one of us\cite{cruz2021testing} for the unitary errors of two-qubit operations implemented in transmon-based quantum hardware, namely a biased-\textsc{cnot} (\textsc{bcnot}) gate. In the current quantum processors developed by IBM~Q, the two-qubit interaction that implements a \textsc{cnot} gate is the so-called cross-resonance (\textsc{cr}) gate. In theory, the \textsc{cr} pulse Hamiltonian should only generate a $ZX$ interaction term, the time evolution of which results in a \textsc{cnot} (up to single-qubit rotations) for an appropriate duration of the dynamics. In practice, however, control errors arise due to the challenging calibration procedure and result in small additional error-terms in the interaction. Focusing only on the two-qubit subspace of the effective \textsc{cr} Hamiltonian and ignoring the entanglement with spectator qubits and external degrees of freedom, the most significant of these error-terms have been identified as $IY$, $IZ$, $IX$, $ZY$ and $ZZ$ \cite{sundaresan2020reducing}. The \textsc{bcnot} gate takes these terms into account with five dimensionless parameters, $\{\beta_j\}_{j = 1}^{5}$, that quantify the bias ratios between the coupling strength of these extra error terms and the desired $ZX$ interaction. Its usefulness in modeling experimental data and improving the understanding of the computational outcomes of these quantum processors has been statistically demonstrated with exhaustive experiments on small circuits.

By replacing all \textsc{cnot} gates by \textsc{bcnot} gates in the circuits we provided for the Fredkin and Toffoli gates, in silico numerical simulations were performed to evaluate the performance of these decompositions in approximating the target unitary, both with and without the ECA procedure. We assumed that a \textsc{bcnot} between each different pair of qubits has different bias parameters. However, these parameters remain fixed over time for a \textsc{cnot} gate applied in the same qubit pair more than once in the circuit, in order to simulate a systematic miscalibration of that gate. The numerical study began by uniformly sampling the five bias ratios $\{\beta_j\}_{j = 1}^{5}$ in the interval $\left[-\beta_{\max},\beta_{\max}\right]$ to assign them to the \textsc{bcnot} model of each qubit pair in the circuit. Having defined all two-qubit gates under the noise model, the unitary representations of the equivalent circuits for the Fredkin (Toffoli) gate were obtained by replacing every \textsc{cnot} appearing in the circuit by the respective \textsc{bcnot}. The diamond distance of each of these unitaries to the target unitary was computed and their average was calculated. The same unitaries were also used to build the corresponding uniformly-mixed-unitary channel (see Eq. \ref{eq:eca_channel}), and the diamond distance from the channel to the target unitary was also computed. The \texttt{Qutip 4.7} open-source software library \cite{johansson2012qutip} was employed to perform these computations through a simplified semi-definite program method \cite{watrous2012simpler}. This procedure was repeated $B$ times, each with a different sampling of the biases in the interval mentioned above for a given $\beta_{\max}$. With the resulting $B$ values for the diamond distances of the channel and the average diamond distances of the unitaries of each circuit, two separate averages and standard deviations were calculated. This process was repeated inside an external loop that varied $\beta_{\max}$ from $0$ to $0.5$.

\begin{figure}[t]
\begin{centering}
\includegraphics[width=1\linewidth]{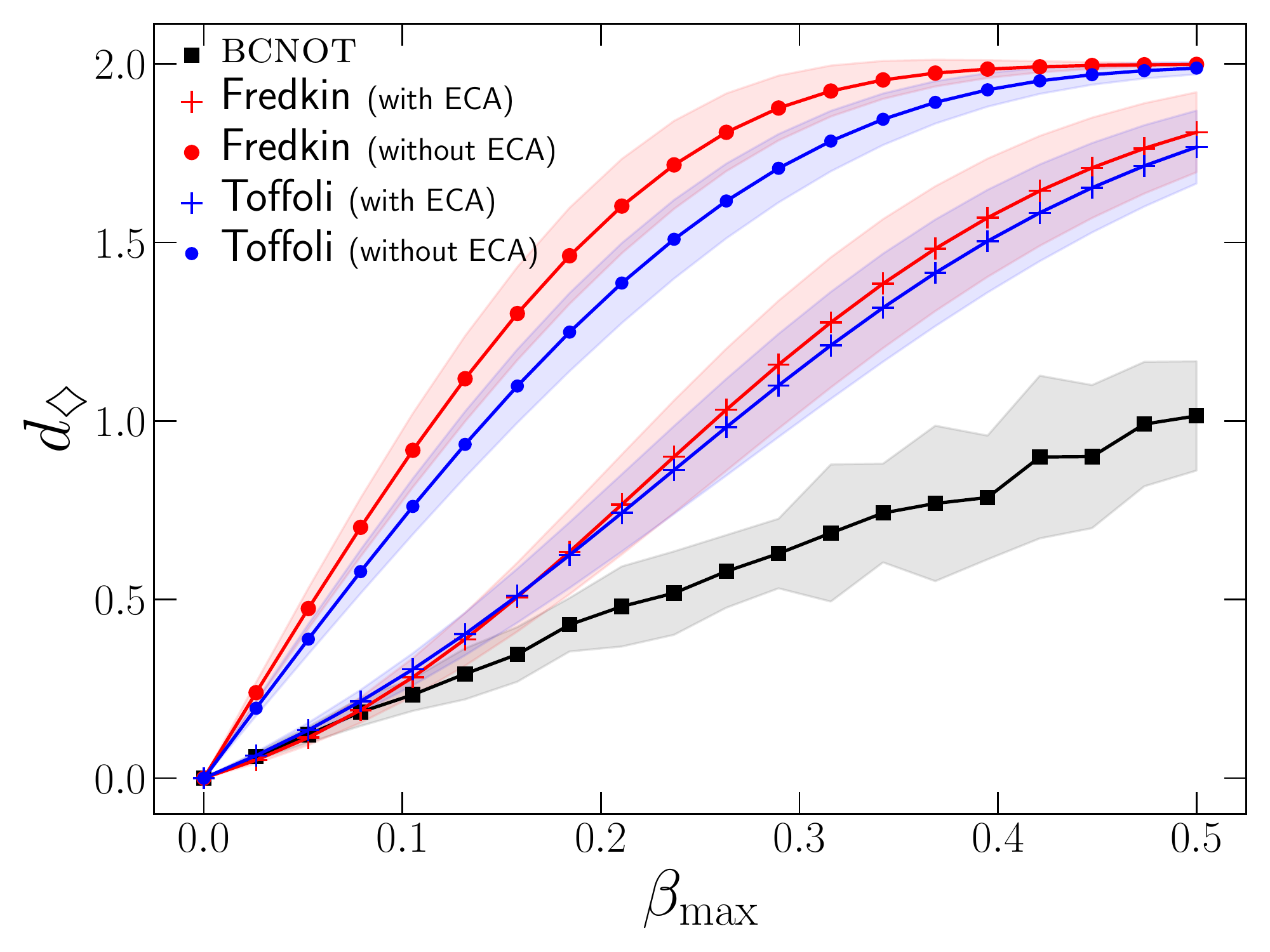}
\par\end{centering}
\caption{Impact of equivalent circuit averaging (ECA) on the approximation of the Fredkin and Toffoli gates using the multiple logically equivalent circuits discussed in Section \ref{sec:adj-dec} subject to a coherent-noise model where every \textsc{cnot} is replaced by a biased-\textsc{cnot} (\textsc{bcnot})\cite{cruz2021testing}. The degree of approximation to the exact unitary is quantified through the diamond distance $d_{\diamondsuit}$, which is plotted against the maximum magnitude $\beta_{\textrm{max}}$ of the bias ratios of the noise model. The numerical simulation procedure is detailed in the main text. For each of the $20$ different values of $\beta_{\textrm{max}}$, $B = 20$ different \mbox{\textsc{bcnot}} models were generated. The diamond distance between the \mbox{\textsc{bcnot}} and the \textsc{cnot} gates is plotted for reference. The width of each shaded region represents two standard deviations. The ECA implementation results in a significant reduction of $d_{\diamondsuit}$ for the Fredkin and Toffoli circuits compared to single-circuit implementations. The systematic difference in $d_{\diamondsuit}$ between the Toffoli and Fredkin circuits, with or without ECA, is due to the Toffoli circuit having one fewer \textsc{cnot} gate, making it less susceptible to the coherent errors.}
\label{fig:eca-diamond}
\end{figure}

The results are plotted in Fig.~\ref{fig:eca-diamond} for a total of $B = 20$ \textsc{bcnot} models generated for each value of $\beta_{\textrm{max}}$. For both the Fredkin (red) and Toffoli (blue) gates, the diamond distance relative to the exact unitary representation of the gate of the uniformly-mixed-unitary channel resulting from the ECA methodology is noticeably lower than the average diamond distance for a single circuit. The black line shows the diamond distance for a single \textsc{bcnot} with respect to the exact \textsc{cnot} for reference. Naturally, the diamond distances of the Toffoli and Fredkin gates are greater than that of the \textsc{bcnot}, as each takes $6$ and $7$ \textsc{cnot}s, respectively, since all-to-all connectivity was assumed. The consistently lower diamond distance for the Toffoli gate relative to the Fredkin gate is due to the extra \textsc{cnot} involved in the decomposition of the latter.

\subsection{Application to quantum simulation: An example}\label{sec:ECA_Hubbard}

As a proof of concept of the application of equivalent circuit averaging to the determination of expectation values of physical quantities in digital quantum simulation, in this section we consider the estimation of the energy of the ground state of the Fermi-Hubbard model \cite{Gutzwiller1963, Kanamori1963, Hubbard1963} on a two-site lattice at half-filling using the Gutzwiller wave function\cite{Gutzwiller1963, murta2021gutzwiller}. 

The Fermi-Hubbard model is a canonical description of strongly-correlated electrons, capturing the competition between the kinetic energy, which favors the delocalization of electrons, and the potential energy, which tends to localize electrons due to the repulsive Coulomb interaction between like charges. Specifically, the electrons are assumed to be in a lattice, where each site represents an orbital of an atom that is part of the crystalline structure of a solid. The hopping of an electron from one site to a nearest-neighboring one lowers the energy by $-t < 0$. Each site can only be occupied by two electrons at most, one with spin-$\uparrow$ and another with spin-$\downarrow$; such a double occupancy of a site imposes an energy penalty of $U >0$. For a sufficiently low temperature, the electrons under the Fermi-Hubbard model take the configuration that minimizes the total energy --- the so-called \textit{ground state}. 

On quantum hardware, adopting the Jordan-Wigner transformation to map electrons to qubits \cite{McArdle2020}, each site is encoded by two qubits, one to store in the computational basis states the number of spin-$\uparrow$ electrons at that site (either $0$ or $1$) and another for spin-$\downarrow$. Here we consider a two-site lattice, so four qubits are required to store the wave function. Assuming half-filling and net zero magnetization --- i.e., there are as many electrons as the number of sites, one with spin-$\uparrow$ and another with spin-$\downarrow$ ---,  the Gutzwiller wave function \cite{Gutzwiller1963} encodes the exact ground state for the two-site case through a suitable choice\footnote{Specifically, $g = 1 - \frac{4t}{U + \sqrt{U^2 + 16t^2}}$, following the definition from Ref.\cite{murta2021gutzwiller}.} of its single free parameter $g$. This ansatz is prepared on quantum hardware following the scheme proposed by one of us \cite{murta2021gutzwiller}. At each site, a controlled-controlled-$R_y$ ($ccR_y$) gate with the two qubits that encode the spin-$\uparrow$ and spin-$\downarrow$ occupations at that site acting as control-qubits and an ancillary qubit initialized in the fiducial state $\ket{0}$ acting as the target-qubit is applied to the ground state of the non-interacting model (i.e., for $\frac{U}{t} = 0$, which is just a Slater determinant\cite{Kivlichan2018, Jiang2018}). The Gutzwiller parameter $g$ sets the angle of the $R_y$ gate\footnote{Concretely, if $\theta$ is the parameter of the $ccR_y$ gate, $\theta(g) = 2 \arctan(\sqrt{2g - g^2}/(1-g))$.}. After applying the $ccR_y$ gate, the ancilla is measured in the computational basis and only the trials that yield the fiducial state $\ket{0}$ are retained, thus resulting in a non-deterministic preparation scheme. The greater $\frac{U}{t}$, the lower the probability of success, converging to $\frac{1}{4}$ as $\frac{U}{t} \to \infty$ for the two-site case. Overall, the $6$-qubit circuit --- i.e., $4$ qubits to encode the ground state and $2$ ancillas, one for each site --- comprises two $ccR_y$ gates, each being decomposed in terms of two Toffoli gates, thus resulting in a total of four Toffoli gates. A scheme of the quantum circuit can be found in Appendix \ref{AppD}.

Having prepared the exact ground state $\ket{\psi_0}$ of the two-site Fermi-Hubbard model for a given set of parameter values $t$ and $U$, its energy is estimated by computing the expectation value of the Fermi-Hubbard Hamiltonian $H$, $\braket{\psi_0 | H | \psi_0}$. Using the Jordan-Wigner transformation and ordering the qubits by spin instead of site (see Appendix \ref{AppD}), the expansion of the Hamiltonian in the Pauli basis is given by
\begin{equation}
    \begin{aligned}
    H = & -\frac{t}{2} \big( X_{0} X_{1} + X_{2} X_{3} + Y_{0} Y_{1} + Y_{2} Y_{3} \big) \; + \\
    & + \frac{U}{4} \big( 2 \times \mathbf{1} - Z_{0} - Z_{1} - Z_{2} - Z_{3} + Z_{0} Z_{2} + Z_{1}Z_{3} \big).
    \end{aligned}
\label{eq:Hamiltonian_sampling}
\end{equation}
There are three sets of commuting terms: $\{ X_{0} X_{1}, X_{2} X_{3} \}$, $\{ Y_{0} Y_{1}, Y_{2} Y_{3} \}$, and $\{ Z_{0}, Z_{1}, Z_{2}, Z_{3}, Z_{0} Z_{2}, Z_{1}Z_{3} \}$. All terms within each set can be measured simultaneously. Computing the expectation value of the Pauli strings in the first set amounts to measuring all four qubits in the main register in the $X$ basis, and similarly for the second and third sets with respect to the $Y$ and $Z$ bases, respectively.  

\begin{figure}[t]
\begin{centering}
\includegraphics[width=\linewidth]{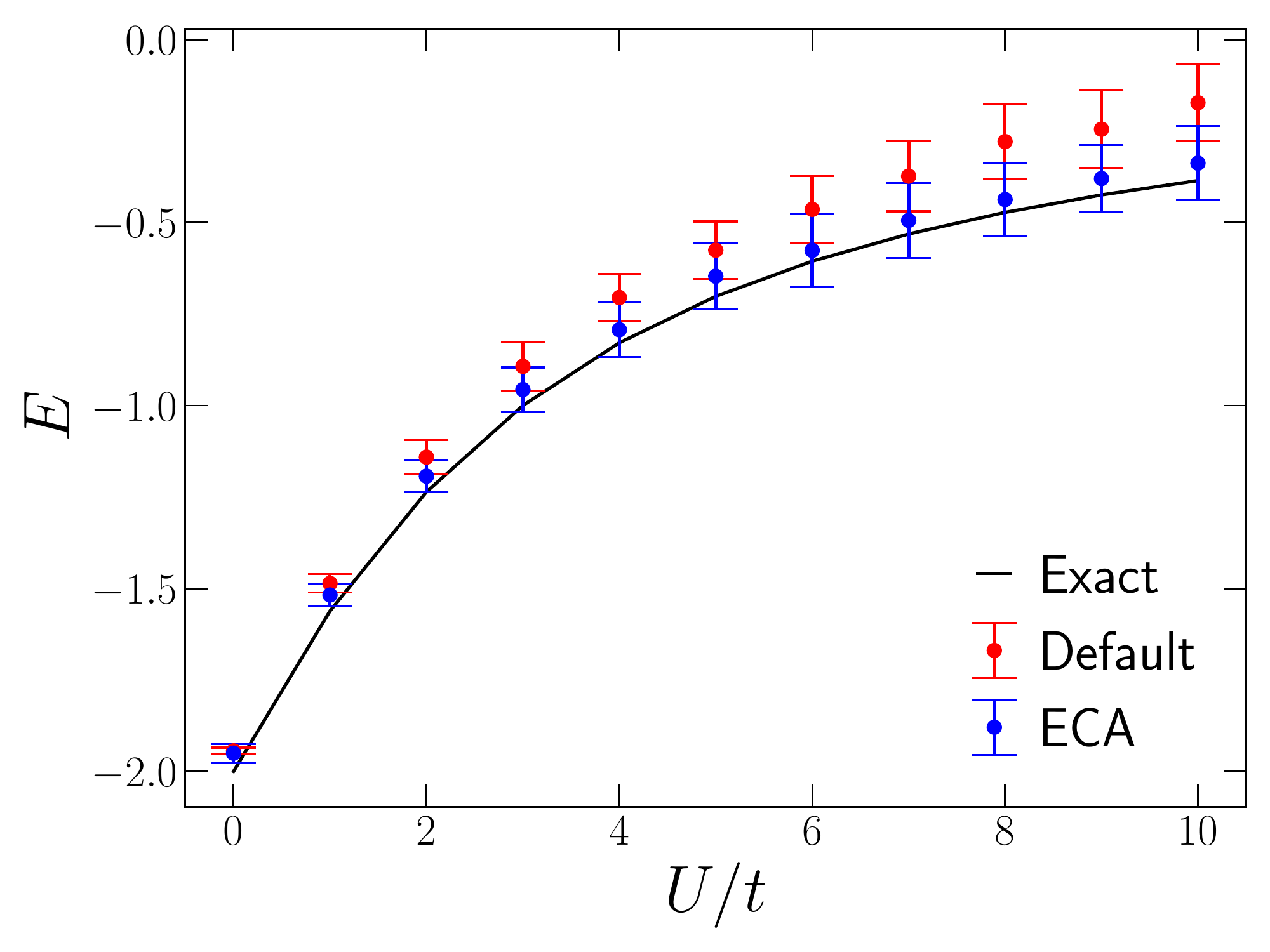}
\par\end{centering}
\caption{Coherent error mitigation via equivalent circuit averaging (ECA) in the ground state energy estimation of the two-site Fermi-Hubbard model. $t$ is the hopping constant and $U$ is the Hubbard parameter of the Fermi-Hubbard model. A \textsc{bcnot} noise model with $\beta_{\textrm{max}} = 0.04$ was considered. The ground state was prepared via the quantum circuit shown in Appendix \ref{AppD}. The exact ground state energy is shown in black. A total of 100,000 samples were generated to estimate each set of commuting terms in the Hamiltonian stated in Eq. (\ref{eq:Hamiltonian_sampling}). These 100,000 samples were divided into 100 trials. For the default approach, the same circuit was employed to prepare the ground state across all trials, replacing each of the four occurrences of the Toffoli gate by the circuit shown inside the blue solid-line box in Fig.~\ref{fig:standard-decomps}(e). The corresponding results are shown in red. For the ECA method, in each of the 100 trials, a new circuit was generated by selecting a circuit at random from the set of $48$ logically equivalent ones for the Toffoli gate with all-to-all connectivity introduced in Section \ref{sec:adj-dec} for each of the four Toffoli gates of the circuit. The respective results are presented in blue. Only the samples for which the Gutzwiller wave function was successfully prepared were considered, due to the non-deterministic nature of the preparation scheme. This contributes to the rise in the size of the error bars as $\frac{U}{t}$ increases, since the probability of success of the preparation scheme decreases with $\frac{U}{t}$ down to a minimum of $\frac{1}{4}$ as $\frac{U}{t} \to \infty$ for the two-site case.}
\label{fig:gutzwiller-sim}
\end{figure}

In order to demonstrate what would be observed in practice, Fig.~\ref{fig:gutzwiller-sim} shows the finite-statistics estimated energy of the ground state $\ket{\psi_0}$ of the two-site Fermi-Hubbard model in the presence of a \textsc{bcnot} coherent-noise model with $\beta_{\textrm{max}} = 0.04$. Other simulations under \textsc{bcnot} noise models with different randomly generated parameters for the same $\beta_{\max}$ were also performed, producing analogous results. All-to-all qubit connectivity is assumed. The exact ground state energy is shown in black for reference. The results presented in red correspond to the default option where the textbook circuit for the Toffoli gate (see circuit inside blue solid-line box in Fig.~\ref{fig:standard-decomps}(e)) was repeated at all four occurrences of the Toffoli gate in the circuit that prepares $\ket{\psi_0}$. The results in blue correspond to the ECA methodology, whereby one of the $48$ logically equivalent circuits generated for the Toffoli gate was sampled at random for each of the four instances the Toffoli gate appears in the circuit. To allow for a fair comparison between the default and ECA approaches, in both cases, for each value of $\frac{U}{t}$, $100$ different sampling trials were carried out, each involving $1000$ measurements. Of the total of 100,000 samples, only those for which both ancillas were measured in the fiducial state $\ket{0}$ --- thus signalling the successful preparation of $\ket{\psi_0}$ in the ideal noiseless scenario --- were used to estimate the energy. This accounts, in part, for the larger error bars observed as $\frac{U}{t}$ increases: fewer trials were used to estimate the energy, so the shot noise is greater. The ECA approach yields estimates of the ground state energy closer to the exact value across the whole range of values of $\frac{U}{t}$, thus handling the effect of the \textsc{bcnot} coherent errors more effectively than the default method. This implementation was not even intended to address the coherent errors introduced by the \mbox{\textsc{bcnot}s} present in the first part of the circuit (see red dashed-line box in Fig.\ref{fig:qc_FHM_gs}) as the averaging over equivalent circuits only considers the second part (see blue solid-line box in Fig.\ref{fig:qc_FHM_gs}) where the four Toffoli gates are present. Nevertheless, of the $28$ \textsc{bcnot}s present in the circuit, $24$ are found in the latter part, so most of the impact of the coherent-noise model is addressed by ECA.

\subsection{Experimental testing}\label{sec:experiment}

Finally, we conducted an experimental evaluation of the equivalent circuit averaging protocol using an IBM~Q quantum processor to validate its performance on a physical device. Specifically, we considered the \textsc{swap} test~\cite{barenco1997stabilization, buhrman2001quantum} with single-qubit states as the application example. The \textsc{swap} test is a quantum algorithm that estimates the fidelity \mbox{$F=\left|\left\langle \psi_{1}|\psi_{2}\right\rangle \right|^{2}$} for two input states $\left|\psi_1\right\rangle $ and $\left|\psi_2\right\rangle $ without performing full tomography of each one separately. For single-qubit states it requires three qubits: one to prepare each input state, and a third auxiliary qubit to be measured in the computational basis to estimate the fidelity from the expectation value \mbox{ $\hat{F} \equiv \left\langle Z\right\rangle = \mathrm{p}(0) - \mathrm{p}(1)$ }. Besides two Hadamard gates, the procedure only employs one Fredkin gate, which can be implemented by making use of our logically equivalent circuit decompositions.

Due to the restricted qubit connectivity of the hardware, we opted to test the \textsc{cswap} decompositions obtained for linear connectivity with the control-qubit at one of the ends, for which we collected equivalent circuits with eight different entangling gate structures (see Table~\ref{tab:stats}). Besides these structural differences, our method also returns variations in single-qubit gates, producing a multitude of different circuits. The count of these variations is not included in Table~\ref{tab:stats} because, as mentioned previously, coherent two-qubit gate errors are more significant than single-qubit ones. On top of that, effective techniques such as randomized compiling can easily add variations to single-qubit gates\cite{wallman2016noise}. Nevertheless, in the experimental implementation, we could leverage all our circuits to increase the diversity of logically equivalent decompositions. Therefore, the 404 circuits with minimal \textsc{cnot}-count that we obtained were transpiled into the native gate set $\mathcal{G}_{\mathrm{IBMQ}}=\left\{ \textsc{cnot}, R_z, \sqrt{X}, X \right\} $ of IBM Q devices and sorted by depth. Since there were more circuits than deemed necessary and their depths (including both CNOTs and single-qubit gates) varied significantly from 28 to 43, a cutoff depth value of 31 was defined and only the shallowest circuits were kept. This value was chosen so that all eight entangling gate structures were represented. In the end, 40 equivalent circuits, with depths of 28 (2 circuits), 29 (7 circuits), 30 (6 circuits), and 31 (25 circuits), were considered. The number of circuits per entangling gate structure was 6, 3, 5, 3, 7, 1, 9, and 6.

\begin{figure}[t]
\begin{centering}
\includegraphics[width=\linewidth]{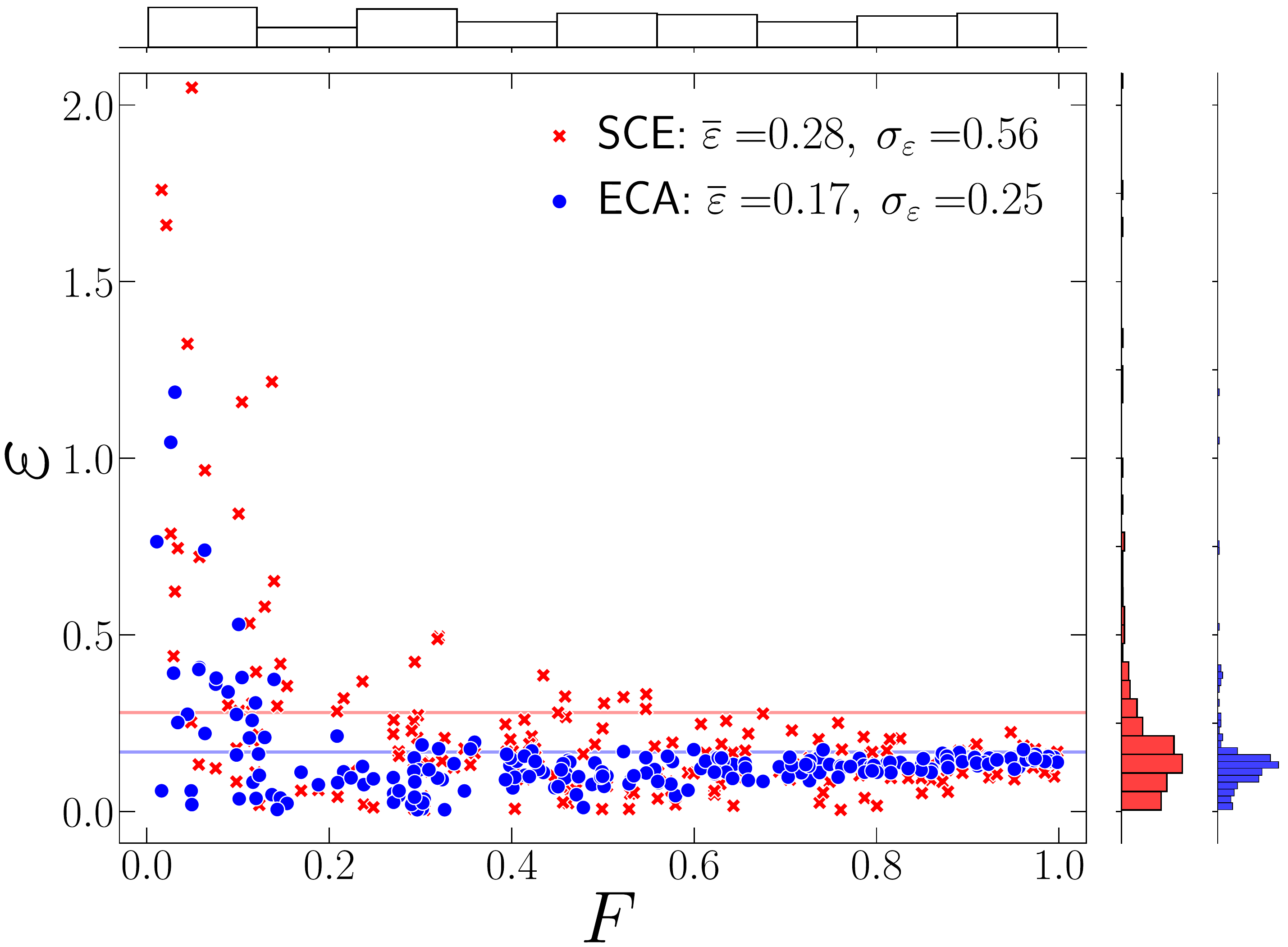}
\par\end{centering}
\caption{Coherent error mitigation via equivalent circuit averaging (ECA) in the \textsc{swap} test of 200 pairs of Haar random single-qubit states performed on the \texttt{ibmq\_lima} quantum processor from IBM Corp. For each pair of states, ECA is compared against a single-circuit execution (SCE) using the relative accuracy error $\epsilon$. The average $\epsilon$ (horizontal colored lines) is significantly reduced from 0.28 to 0.17 when the ECA protocol is applied to combine the shot statistics of different circuit decompositions. The standard deviation of the errors is also noticeably reduced when ECA is performed, from 0.56 to 0.25, improving the predictability of results. The histograms of the marginal distributions are also displayed. Tests were uniformly performed for fidelity, $F$, in the range from 0.01 to 1.}
\label{fig:swaptest-experiment}
\end{figure}

The experimental protocol started by defining 200 pairs of Haar random single-qubit states, generated by applying a Haar random unitary to a fixed pure state, and keeping only the pairs with $F>0.01$. Then, with the aim of estimating the fidelity of each pair of states, an independent experiment was prepared for each of them to be carried out in two different protocols: as a single-circuit execution (SCE) or with ECA. A budget of $S=980,000$ shots was given to each protocol. For the SCE protocol, one of the 40 equivalent Fredkin decompositions was sampled\footnote{Note that a different decomposition was sampled for each pair of states, i.e. each different experiment.} and the complete circuit for the \textsc{swap} test was put together by initializing it with the gate sequence that prepared each input state \cite{shende2006synthesis}. This circuit was run $S$ times and $\hat{F}$ was estimated from the measurement outcomes. For the ECA protocol, an equal share of $s=S/8$ shots was given to each entangling gate structure, where the $s$ circuits to be used were defined by sampling from the Fredkin decompositions with the entangling gate structure under consideration. The initial states were prepared with the same algorithm as before, and the resulting $S$ shots were combined to compute $\hat{F}$. All the circuits and shots in one experiment --- comprising both protocols --- were executed within the same \emph{job} in the \texttt{ibmq\_lima} processor\footnote{The \texttt{ibmq\_lima} is a Falcon r4T processor with Quantum Volume 8 made available for free of charge use by IBM~Q, with no special credentials required.} to assure a fair comparisson under the same experimental conditions. Besides the circuits under evaluation (49 copies of only one circuit for the single-circuit execution protocol, and 49 logically equivalent circuits for the ECA protocol) each job included 2 additional circuits to calibrate the measurement error mitigation protocol~\cite{QiskitMEM}, which was also tested with and without combining it with our ECA error mitigation method. Within each job, the first shot of each circuit was executed sequentially before moving on to the second shot, and so on until all of the circuits in the job ran for $20,000$ shots each, totaling $980,000$ shots for each protocol. With 200 pairs of random states and $\sim9$ minutes per job, the full runtime of all the independent experiments was of around 30 hours, spread over two days\footnote{It is worth noting that \texttt{ibmq\_lima} may have undergone recalibration during this time. However, all executions evaluating the performance of the \textsc{swap} test for one pair of random states, using both protocols, were conducted within a single, uninterrupted job lasting approximately 9 minutes. Hence, analyzing the results of all experiments together remains valid even if the device undergoes recalibration between jobs. Our primary focus is to assess whether the performance with ECA improves relative to the single-circuit within each job, independently.}.

Having completed all the experiments, our analysis started by comparing the values of the measured fidelity $\hat{F}$ and the expected value $F$, revealing an anticipated behavior: In both protocols, in the cases where $F \approx 0$, the value of the estimated value $\hat{F}$ tends to be slightly higher than the theoretical value $F$, as it becomes more challenging to reduce the error further due to random errors during circuit execution; conversely, when $F \approx 1$, any error introduced tends to decrease $\hat{F}$, emphasizing the sensitivity of fidelity estimation to errors in such scenarios. More comprehensively, Fig.~\ref{fig:swaptest-experiment} presents the evaluation of the relative accuracy error $\epsilon = \left| \hat{F} - F \right| / F$ in fidelity estimation with and without the ECA protocol. The plot showcases the remarkable improvements achieved through the application of ECA. It is evident that the ECA protocol not only reduces errors but also diminishes their variability. To quantify this observation, we computed the average ($\overline{\epsilon}$) and standard deviation ($\sigma_{\epsilon}$) of all relative errors. ECA proves to be highly effective in error reduction when compared to SCE, reducing the average relative error from 28\% to 17\%, and their standard deviation from 56\% to 25\%.

While Fig.~\ref{fig:swaptest-experiment} exclusively displays the results of our ECA method, it is important to note that it can seamlessly be complemented with other error mitigation protocols, such as measurement error mitigation (MEM)~\cite{QiskitMEM, Nation2021}. Although not displayed in the figure, we compared ECA with MEM~\cite{QiskitMEM} to further benchmark our method. Specifically, MEM applied to the SCE protocol yields a non-significant 1\% improvement in $\overline{\epsilon}$, reducing it from 28\% to 27\%, with the associated $\sigma_{\epsilon}$
actually increasing to 62\%. In stark contrast, ECA significantly enhances the results, reducing the average relative error from 28\% to 17\%, as mentioned above. Moreover, we coupled ECA with MEM, demonstrating its potential for even greater error reduction. We observed that when MEM is coupled with ECA, it achieves the best performance, with an average relative error of only $\overline{\epsilon}=16\%$ with standard deviation also reaching $\sigma_{\epsilon}=16\%$.

In addition to supplementing ECA with MEM, it might be worth considering pairing it with a compatible technique for mitigating incoherent errors. Because these errors are expected to occur randomly and independently of the circuit decomposition, and because their level should be similar in circuits with comparable depths, ECA should have no impact on them. Therefore, complementing ECA with incoherent error mitigation might improve fidelity further.

\section{Conclusion \label{sec:concl}}

The Fredkin and Toffoli gates play a prominent role in quantum computing, underscoring the critical importance of efficiently decomposing these three-qubit gates in terms of \textsc{cnot}s and single-qubit gates. In this paper, we have provided multiple decompositions of the Fredkin and Toffoli gates that achieve, to the best of our knowledge, an optimal \textsc{cnot} count, thus being relevant for near-term quantum hardware. The savings in \textsc{cnot} count produced by our ZX-calculus-based optimization scheme were especially pronounced under qubit connectivity constraints. Since the generation of the multiple equivalent quantum circuits herein presented demanded a considerable amount of time of computation, these circuits have been stored in memory to be retrieved when required.

Besides considering the case where the three qubits on which the Toffoli and Fredkin gates act nontrivially are adjacent, we have also explored the scenario where they are separated from one another in an architecture subject to connectivity constraints. In particular, we have devised an improved scheme to efficiently reroute the qubits of long-range Fredkin and Toffoli gates by replacing a \textsc{swap} gate with a \mbox{\textsc{cnot-swap}}. Although it only successfully swaps one of the qubits while leaving the other one dirty, it takes only two \textsc{cnot}s as opposed to the three required by a perfect \textsc{swap}. We employed this \textsc{cnot-swap}-based rerouting scheme to bring the three active qubits next to one another in order to apply our local Toffoli or Fredkin gate decompositions before returning them back to their original positions whilst ensuring that the idle qubits are left in their starting state. Consequently, the \textsc{cnot} count and depth for implementing these three-qubit gates was further reduced.

The use of \textsc{cnot-swap}s is not restricted to the implementation of the Fredkin and Toffoli gates. In fact, the replacement of the standard \textsc{swap} with a \textsc{cnot-swap} --- thus saving one \textsc{cnot} for every substitution --- applies generally to the rerouting of the control-qubits of any multi-controlled-gate and of the target-qubit of any multi-controlled-\textsc{not} operation, as well as any qubit from a multi-qubit gate with respect to which the matrix representation of the gate is diagonal in the computational basis. A noteworthy example of application of this general scheme corresponds to the implementation of a long-range \textsc{cnot} --- i.e., a \textsc{cnot} between two qubits that are not directly connected to each other. In addition to yielding the optimal \textsc{cnot} count decomposition of the long-range \textsc{cnot} when there are $n=1$ or $n=2$ idle qubits between the active ones --- as confirmed by an exhaustive search with circuits comprising only \textsc{cnot}s ---, this \textsc{cnot-swap}-based decomposition results in exactly $4n$ \textsc{cnot} gates and depth a of $\sim n$. Although this \textsc{cnot} count scaling had already been achieved by Shende \textit{et al.} \cite{shende2006synthesis}, their decomposition did not offer the possibility of compressing circuit depth, thus being restricted to a depth scaling of $4n$ as well. Our \textsc{cnot-swap} methodology, in turn, does allow for the parallelization of \textsc{cnot}s by moving both the control and the target qubits towards each other simultaneously and by permuting commuting \textsc{cnot}s in the rerouting layers, as illustrated in Fig.~\ref{fig:long-cnot}(c). The \textsc{cnot-swap} decomposition of the long-range \textsc{cnot} therefore combines the best of both worlds.

Having multiple logically equivalent circuits with different entangling gate structures that realize the Toffoli and Fredkin gates opens a number of possibilities for overall circuit optimization and error mitigation. In this regard, we have explored the use of equivalent circuit averaging (ECA) --- i.e., combining the measurement statistics of multiple different but logically equivalent circuits as opposed to repeating the same circuit multiple times --- to address the effects of coherent noise sources. Using a realistic coherent-noise model that accounts for the leading-order biases in the implementation of the \textsc{cnot} via the cross-resonance gate in transmon-based quantum hardware, the uniformly-mixed-unitary channel resulting from the ECA methodology was shown to approximate the exact Fredkin and Toffoli unitaries more closely than an average individual circuit by computing the diamond distance. In addition, to illustrate the application of ECA to digital quantum simulation, we employed this methodology in the estimation of the energy of the ground state of the Fermi-Hubbard dimer, having obtained improved results relative to the bare approach using the same coherent-noise model considered in the calculation of the diamond distance. Finally, to confirm the effectiveness of the ECA methodology on actual quantum hardware, an experiment that involved estimating the fidelity between pairs of single-qubit states via the \textsc{swap} test was carried out on an IBM Q processor. ECA was found to reduce both the average relative accuracy error and its variance with respect to the single-circuit approach. The integration of ECA with measurement error mitigation resulted in a further reduction of the average error.

The various decompositions of the Fredkin and Toffoli gates should find wide use in near-term quantum computing hardware. We expect them to be especially useful in solid-state platforms based on superconducting circuits and silicon quantum dots, given the prevalence of qubit connectivity constraints in such cases. Nevertheless, even quantum computing platforms based on trapped ions and cold atoms may benefit from the multiple realizations of the Fredkin and Toffoli gates that assume all-to-all connectivity, namely to perform equivalent circuit averaging to mitigate coherent errors or to unlock opportunities for overall circuit simplifications. While the results presented in Section~\ref{sec:ECA} regarding the implementation of equivalent circuit averaging on quantum processors based on superconducting circuits are promising, further studies involving alternative technological realizations of quantum computers are encouraged.

\begin{acknowledgments}
We acknowledge fruitful discussions with J. P. Pedroso, J. Fern\'andez-Rossier, D. Farina, and E. B\"aumer. P.M.Q.C. acknowledges Funda\c{c}\~{a}o para a Ci\^{e}ncia e a Tecnologia (FCT) Portugal for Grant No. SFRH/BD/150708/2020, the Government of Spain (Severo Ochoa CEX2019-00910-S and TRANQI), Fundació Cellex, Fundació Mir-Puig, Generalitat de Catalunya (CERCA program), and the AXA Chair in Quantum Information Science. B.M. acknowledges support from FCT Grant No. SFRH/BD/08444/2020. Both authors acknowledge use of the IBM Q for this work. The views expressed are those of the authors and do not reflect the official policy or position of IBM Corp. or the IBM Q team.
\end{acknowledgments}

\section*{Data Availability Statement}
The \texttt{OpenQASM} instruction files for the quantum circuits used in this study are openly available\cite{cruz2023Fredkin} in Zenodo at \url{https://doi.org/10.5281/zenodo.10047422}, reference number 10.5281/zenodo.10047422. Access to these files is granted under a CC BY-NC 4.0 license to mitigate the computational cost and time required by quantum compilers to re-obtain them at runtime when executing circuits that make use of these primitives.

\appendix

\section{Controlling a gate with symmetric decomposition \label{AppA}}

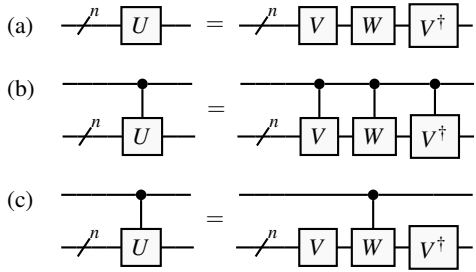
\begin{figure}[h]
\begin{tabular}{l}
(a) \raisebox{-0.4\totalheight}{
\begin{tikzpicture}
\node[scale=1]{
\begin{tikzcd}[row sep={0.7cm,between origins}, column sep=0.2cm]
& \qw & \qwbundle{n} & \qw & \gate{U} & \qw & \qw \midstick[3,brackets=none]{$=$} & \qw & \qwbundle{n} & \qw & \gate{V} & \gate{W} & \gate{V^\dagger} & \qw
\end{tikzcd}
};
\end{tikzpicture}
} \\
(b) \raisebox{-0.7\totalheight}{
\begin{tikzpicture}
\node[scale=1]{
\begin{tikzcd}[row sep={0.7cm,between origins}, column sep=0.2cm]
& \qw  & \qw & \qw & \ctrl{1} & \qw & \qw \midstick[3,brackets=none]{$=$} & \qw & \qw & \qw & \ctrl{1} & \ctrl{1} & \ctrl{1} & \qw \\
& \qw & \qwbundle{n} & \qw & \gate{U} & \qw & \qw & \qw & \qwbundle{n} & \qw & \gate{V} & \gate{W} & \gate{V^\dagger} & \qw
\end{tikzcd}
};
\end{tikzpicture}
} \\
(c) \raisebox{-0.7\totalheight}{
\begin{tikzpicture}
\node[scale=1]{
\begin{tikzcd}[row sep={0.7cm,between origins}, column sep=0.2cm]
& \qw  & \qw & \qw & \ctrl{1} & \qw & \qw \midstick[3,brackets=none]{$=$} & \qw & \qw & \qw & \qw & \ctrl{1} & \qw & \qw \\
& \qw & \qwbundle{n} & \qw & \gate{U} & \qw & \qw & \qw & \qwbundle{n} & \qw & \gate{V} & \gate{W} & \gate{V^\dagger} & \qw
\end{tikzcd}
};
\end{tikzpicture}
}
\end{tabular}
\caption{(a) Symmetric decomposition of $n$-qubit gate $U = V^{\dagger} W V$. (b) Na\"{i}f approach to controlling $U$ adds a control to every element of the circuit. (c) Thanks to the symmetric decomposition, a control at the central gate $W$ suffices to yield the controlled-$U$ gate.}
\label{fig:symm_control}
\end{figure}

Let $U$ be a $n$-qubit gate, for which a symmetric decomposition $U = V^{\dagger} W V$ can be found for some $n$-qubit gates $V$ and $W$. Suppose we wish to add a control-qubit to $U$. We will show that, in general, the controls on $V$ and $V^{\dagger}$ assumed in the na\"{i}f approach illustrated in Fig.~\ref{fig:symm_control}(b) can be skipped. It suffices to control the central gate $W$ (see Fig.~\ref{fig:symm_control}(c)). 

The matrix representation of the left-hand side of the equality stated in Fig.~\ref{fig:symm_control}(c) is
\begin{equation}
    \begin{pmatrix}
    \mathds{1} & 0 \\
    0 & U
    \end{pmatrix}.
\label{eq:LHS}
\end{equation}
The right-hand side of the same equality is given by
\begin{equation}
    \begin{pmatrix}
    V^{\dagger} & 0 \\
    0 & V^{\dagger}
    \end{pmatrix} \;
    \begin{pmatrix}
    \mathds{1} & 0 \\
    0 & W
    \end{pmatrix} \;
    \begin{pmatrix}
    V & 0 \\
    0 & V
    \end{pmatrix}, 
\end{equation}
and, upon matrix multiplication, this results into
\begin{equation}
    \begin{pmatrix}
    V^{\dagger} V & 0 \\
    0 & V^{\dagger} W V
    \end{pmatrix},
\end{equation}
which simplifies to Eq.~(\ref{eq:LHS}) using the unitarity of $V$ and the given symmetric decomposition of $U$.

\section{Changing target-qubit of Toffoli gate \label{AppB}}

\begin{figure}[t]
\centering
\begin{tabular}{lll}
(a) \raisebox{-0.76\totalheight}{
\begin{tikzpicture}
\node[scale=0.7]{
\begin{tikzcd}[row sep={0.9cm,between origins}, column sep=0.2cm]
& \ctrl{1} & \qw \midstick[3,brackets=none]{$=$} & \qw & \ctrl{1} & \qw & \qw \\
& \targ{} & \qw & \gate{H} & \ctrl{1} & \gate{H} & \qw \\
& \ctrl{-1} & \qw & \gate{H} & \targ{} & \gate{H} & \qw 
\end{tikzcd}
};
\end{tikzpicture}
} &  & (b) \raisebox{-0.76\totalheight}{
\begin{tikzpicture}
\node[scale=0.7]{
\begin{tikzcd}[row sep={0.7cm,between origins}, column sep=0.2cm]
& \targ{} & \qw \midstick[3,brackets=none]{$=$} & \qw & \gate{H} & \ctrl{1} & \gate{H} & \qw \\
& \ctrl{-1} & \qw & \qw & \gate{H} & \targ{} & \gate{H} & \qw
\end{tikzcd}
};
\end{tikzpicture}
}\tabularnewline
\multicolumn{3}{l}{
(c) \raisebox{-0.76\totalheight}{
\begin{tikzpicture}
\node[scale=0.7]{
\begin{tikzcd}[row sep={0.7cm,between origins}, column sep=0.2cm]
& \qwbundle{m_1} & \ctrl{1} & \qw & \qw \midstick[5,brackets=none]{$=$} & \qwbundle{m_1} & \ctrl{1} & \qw & \qw \midstick[5,brackets=none]{$=$} & \qwbundle{m_1} & \ctrl{1} & \qw \\
& \gate{H} & \targ{} & \gate{H} & \qw & \qw & \ctrl{1} & \qw & \qw & \qw & \ctrl{1} & \qw \\
& \qwbundle{m_2} & \ctrl{-1} & \qw & \qw & \qwbundle{m_2} & \ctrl{1} & \qw & \qw & \qwbundle{m_2} & \ctrl{1} & \qw \\
& \gate{H} & \ctrl{-1} & \gate{H} & \qw & \gate{H} & \ctrl{1} & \gate{H} & \qw & \qw & \targ{} & \qw \\
& \qwbundle{m_3} & \ctrl{-1} & \qw & \qw & \qwbundle{m_3} & \ctrl{-1} & \qw & \qw & \qwbundle{m_3} & \ctrl{-1} & \qw
\end{tikzcd}
};
\end{tikzpicture}
}}\tabularnewline
\end{tabular}
\caption{(a) Changing target-qubit of Toffoli gate by applying a pair of Hadamard gates, one on either side of the Toffoli, at the old and new target-qubits. (b) Equivalent two-qubit circuit identity reverses direction of \textsc{cnot}. (c) Generalization to arbitary number $n = m_1 + m_2 + m_3 + 1$ control-qubits for multi-controlled-Toffoli gate.}
\label{fig:toffoli_control}
\end{figure}
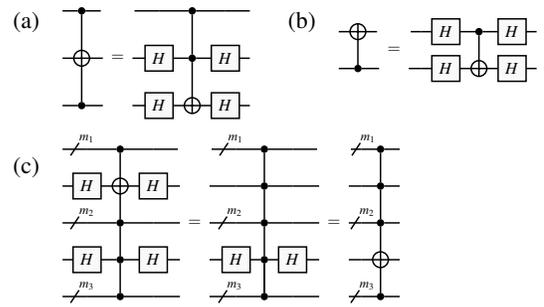

By applying an Hadamard gate on either side of a Toffoli at the target-qubit and a control-qubit, their roles are reversed, i.e., the target-qubit becomes a control-qubit and vice-versa (see Fig.~\ref{fig:toffoli_control}(a)). This result follows from the well-known identity $HXH = Z$ and the fact that the controlled-controlled-\textsc{z} gate is invariant under any permutation of the three qubits on which it acts nontrivially. 

This result is a generalization to three qubits of the more familiar two-qubit result shown in Fig.~\ref{fig:toffoli_control}(b), where the direction of a \textsc{cnot} gate is reversed by applying a pair of Hadamard gates on both qubits, one on either side of the \textsc{cnot}. As shown in Fig.~\ref{fig:toffoli_control}(c), this result is valid for an arbitrary number of control-qubits: A multi-controlled-Toffoli (\textsc{mcx}) gate can always be turned into a multi-controlled-\textsc{z} (\textsc{mcz}) gate by applying the pair of Hadamard gates at the target-qubit, and then a \textsc{mcx} gate with a different target-qubit can be generated by applying another pair of Hadamard gates to the \textsc{mcz} at the new target-qubit. 

It should be stressed, however, that this result is only valid for a single target-qubit, i.e., applying two or more pairs of Hadamard gates to a \textsc{mcz} gate on as many different qubits does not result in a multi-controlled operation with conditional \textsc{not} gates at those qubits.

\section{Two important examples of simplifications of quantum circuits with CNOT-SWAP networks \label{AppC}}

Here, we demonstrate how \textsc{cnot-swap}s can be leveraged to reduce the depth and \textsc{cnot} count of important examples of circuits under linear connectivity constraints. First, the long-range \textsc{cnot} decompositions based on the cnot-swapping methodology (see Section~\ref{sec:nonadj-dec}) are shown to simplify circuits involving sequences of \textsc{cnot} gates with the same control-qubit but different target-qubits, which are commonly found in error correction codes \cite{fowler2012surface, devitt2013quantum}. Then, \textsc{cnot-swap}s are also applied to the circuits that realize complex exponentials of Pauli strings, which are pervasive in quantum simulation\cite{Tacchino2020}.

\subsection{Sequences of CNOTs with shared control-qubit}

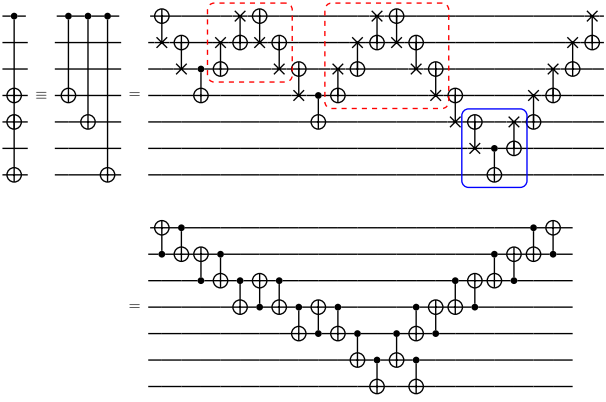
\begin{figure}[t]
\flushleft
\begin{tikzpicture}
\node[scale=0.64]{
\begin{tikzcd}[row sep={0.55cm,between origins}, column sep=0.08cm]
& \ctrl{6} & \qw \midstick[7,brackets=none]{$\equiv$} & \ctrl{3} & \ctrl{4} & \ctrl{6} & \qw \midstick[7,brackets=none]{$=$} & \targ{}   & \qw       & \qw      & \qw \gategroup[3,steps=4,style={dashed, rounded corners, draw=red, inner xsep=0pt, inner ysep=0pt}]{} & \swap{1}  & \targ{}   & \qw       & \qw       & \qw      & \qw \gategroup[4,steps=6,style={dashed, rounded corners, draw=red, inner xsep=0pt, inner ysep=0pt}]{} & \qw      & \swap{1} & \targ{}   & \qw       & \qw       & \qw       & \qw       & \qw      & \qw      & \qw      & \qw      & \qw      & \swap{1} & \qw & \\
& \qw      & \qw & \qw      & \qw      & \qw      & \qw & \swap{-1} & \targ{}   & \qw      & \swap{1} & \targ{}   & \swap{-1} & \targ{}   & \qw       & \qw      & \qw      & \swap{1} & \targ{}  & \swap{-1} & \targ{}   & \qw       & \qw       & \qw       & \qw      & \qw      & \qw      & \qw      & \swap{1} & \targ{}  & \qw & \\
& \qw      & \qw & \qw      & \qw      & \qw      & \qw & \qw       & \swap{-1} & \ctrl{1} & \targ{}  & \qw       & \qw       & \swap{-1} & \targ{}   & \qw      & \swap{1} & \targ{}  & \qw      & \qw       & \swap{-1} & \targ{}   & \qw       & \qw       & \qw      & \qw      & \qw      & \swap{1} & \targ{}  & \qw      & \qw & \\
& \targ{}  & \qw & \targ{}  & \qw      & \qw      & \qw & \qw       & \qw       & \targ{}  & \qw      & \qw       & \qw       & \qw       & \swap{-1} & \ctrl{1} & \targ{}  & \qw      & \qw      & \qw       & \qw       & \swap{-1} & \targ{}   & \qw       & \qw      & \qw      & \swap{1} & \targ{}  & \qw      & \qw      & \qw & \\
& \targ{}  & \qw & \qw      & \targ{}  & \qw      & \qw & \qw       & \qw       & \qw      & \qw      & \qw       & \qw       & \qw       & \qw       & \targ{}  & \qw      & \qw      & \qw      & \qw       & \qw       & \qw       & \swap{-1} & \targ{} \gategroup[3,steps=3,style={solid, rounded corners, draw=blue, inner xsep=0pt, inner ysep=0pt}]{} & \qw      & \swap{1} & \targ{}  & \qw      & \qw      & \qw      & \qw & \\
& \qw      & \qw & \qw      & \qw      & \qw      & \qw & \qw       & \qw       & \qw      & \qw      & \qw       & \qw       & \qw       & \qw       & \qw      & \qw      & \qw      & \qw      & \qw       & \qw       & \qw       & \qw       & \swap{-1} & \ctrl{1} & \targ{}  & \qw      & \qw      & \qw      & \qw      & \qw & \\
& \targ{}  & \qw & \qw      & \qw      & \targ{}  & \qw & \qw       & \qw       & \qw      & \qw      & \qw       & \qw       & \qw       & \qw       & \qw      & \qw      & \qw      & \qw      & \qw       & \qw       & \qw       & \qw       & \qw       & \targ{}  & \qw      & \qw      & \qw      & \qw      & \qw      & \qw & \\
&          & \\    
&          &     &          &          &          & \midstick[7,brackets=none]{$=$}
& \targ{}   & \ctrl{1}       & \qw      & \qw      & \qw      & \qw      & \qw      & \qw      & \qw      & \qw      & \qw & \qw & \qw & \qw & \qw & \qw & \qw & \qw & \qw & \ctrl{1} & \targ{} & \qw \\
&          &     &          &          &          &     & \ctrl{-1} & \targ{}   & \targ{}      & \ctrl{1}      & \qw      & \qw      & \qw      & \qw      & \qw      & \qw & \qw  & \qw & \qw & \qw & \qw & \qw & \qw & \ctrl{1} & \targ{} & \targ{} & \ctrl{-1} & \qw \\
&          &     &          &          &          &     & \qw       & \qw & \ctrl{-1} & \targ{}      & \ctrl{1}      & \targ{}      & \ctrl{1}      & \qw      & \qw & \qw  & \qw      & \qw & \qw & \qw & \qw & \ctrl{1} & \targ{} & \targ{} & \ctrl{-1} & \qw & \qw & \qw \\
&          &     &          &          &          &     & \qw       & \qw       & \qw  & \qw & \targ{}      & \ctrl{-1}      & \targ{}      & \ctrl{1} & \targ{}  & \ctrl{1}      & \qw      & \qw & \qw & \ctrl{1} & \targ{} & \targ{} & \ctrl{-1} & \qw & \qw & \qw & \qw & \qw \\
&          &     &          &          &          &     & \qw       & \qw       & \qw      & \qw  & \qw & \qw      & \qw & \targ{} & \ctrl{-1}      & \targ{}     & \ctrl{1}     & \qw & \ctrl{1} & \targ{} & \ctrl{-1} & \qw & \qw & \qw & \qw & \qw & \qw & \qw \\
&          &     &          &          &          &     & \qw       & \qw       & \qw      & \qw & \qw  & \qw & \qw  & \qw      & \qw      & \qw      & \targ{}      & \ctrl{1} & \targ{}  & \ctrl{1} & \qw & \qw & \qw & \qw & \qw & \qw & \qw & \qw \\
&          &     &          &          &          &     & \qw       & \qw       & \qw      & \qw  & \qw      & \qw  & \qw      & \qw      & \qw      & \qw      & \qw      & \targ{} & \qw  & \targ{} & \qw & \qw & \qw & \qw & \qw & \qw & \qw & \qw
\end{tikzcd}
};
\end{tikzpicture}
\caption{Example of a single-control multi-target-\textsc{not} gate decomposition in terms of nearest-neighbor \textsc{cnot} gates. The circuit is simplified by making use of the \textsc{cnot-swap} decomposition of long-range \textsc{cnot}s (see Fig.~\ref{fig:long-cnot}) and eliminating conjugated pairs of \textsc{cnot-swap} gates acting on the same qubits when possible, as highlighted inside the red dashed-line boxes. The subcircuit in the blue solid-line box is further simplified with the optimal decomposition of a \textsc{cnot} with an idle qubit between the control- and target-qubits. The final decomposition comprises $22$ \textsc{cnot} gates and has depth $21$.}
\label{fig:SCMT-NOT}
\end{figure}

Fig.~\ref{fig:SCMT-NOT} shows an example of a quantum circuit with three consecutive \textsc{cnot} gates that share the same control-qubit but act on different target-qubits. As the leftmost scheme suggests, such a sequence of \textsc{cnot}s can be regarded as a single-control-multi-target-\textsc{not} gate. Assuming linear qubit connectivity, each long-range \textsc{cnot} is implemented by moving the control-qubit via \textsc{cnot-swap}s until it is next to the target-qubit, applying a \textsc{cnot} gate, and returning the control-qubit to its original position via \textsc{cnot-swap}s. The \textsc{cnot-swap}s within the red dashed-line boxes highlighted in the scheme after the second equality of Fig.~\ref{fig:SCMT-NOT} cancel out in pairs, which greatly reduces the \textsc{cnot} count and depth. Finally, the subcircuit within the blue solid-line box, which would take $5$ \textsc{cnot}s upon decomposing the \textsc{cnot-swap}s, can be replaced by the $4$-\textsc{cnot} circuit shown in the dashed-line box of Fig.~\ref{fig:long-cnot}. All in all, the full circuit has a total of $22$ \textsc{cnot}s and depth $21$. 

Had we implemented each of the three long-range \textsc{cnot}s via the method first introduced by Shende \textit{et al.} \cite{shende2006synthesis}, we would have obtained a \textsc{cnot} count of $30$ and depth $29$. Like the \textsc{cnot-swap}-based approach described in Fig.~\ref{fig:SCMT-NOT}, the standard approach of moving the control-qubit via conventional \textsc{swap}s also allows for the cancellation of many gates, resulting in a \textsc{cnot} count and depth of $29$. Alternatively, we can make use of the \textsc{cnot-swap} decomposition of the long-range \textsc{cnot}s whilst moving both the control- and target-qubits in parallel towards each other; compared to the case where only the control-qubit is moved (see Fig.~\ref{fig:SCMT-NOT}), the depth is reduced from $21$ to $19$, but the \textsc{cnot} count increases from $22$ to $31$, as fewer pairs of \textsc{cnot-swap}s cancel out. 

Although the \textsc{cnot-swap}-based methods herein introduced result in a shallower circuit for the example considered in Fig.~\ref{fig:SCMT-NOT}, we note that this advantage relative to the long-range \textsc{cnot} decomposition of Shende \textit{et al.} \cite{shende2006synthesis} may not be observed for all circuits with successive \textsc{cnot}s sharing the same control-qubit. In fact, in the cases where all target-qubits are adjacent to one another (though distant from the shared control-qubit), the method by Shende \textit{et al.} \cite{shende2006synthesis} achieves a lower \textsc{cnot} count after straightforward simplifications of the global circuit. For example, if the target-qubits of the three \textsc{cnot}s in Fig.~\ref{fig:SCMT-NOT} were the three bottommost qubits in the scheme, the \textsc{cnot-swap} method would result in $20$ \textsc{cnot}s and depth $18$, while the long-range \textsc{cnot} method due to Shende \textit{et al.}\cite{shende2006synthesis} would produce a circuit with $16$ \textsc{cnot}s and depth $14$. The advantage of one long-range \textsc{cnot} decomposition over the other for the overall simplification of these circuits depends on the specific \textsc{cnot} sequence under consideration, the number of qubits involved, and the adjacency relations between all target-qubits. In practice, a compilation procedure could be implemented to choose the combination of different long-range \textsc{cnot} decompositions that yields the shallowest circuit.

\subsection{Complex exponentials of Pauli strings}

Let $P$ be a $n$-qubit Pauli string, i.e., $P \in G_n \equiv \{\mathbb{1}_{2 \times 2}, X, Y, Z \}^{\otimes n}$, where $G_n$ is the Pauli group on $n$ qubits \cite{nielsen2002quantum}. Any unitary of the form $e^{-i \theta P}$ with $\theta \in \mathbb{R}$ can be implemented with $2 (s-1)$ \textsc{cnot}s, where $s \leq n$ is the number of qubits on which $P$ acts nontrivially (i.e., the number of occurrences of $X$, $Y$ or $Z$ in the Pauli string $P$, with the remaining $n-s$ elements of the tensor product corresponding to $\mathbb{1}_{2 \times 2}$). The key idea\cite{nielsen2002quantum} behind this decomposition is the fact that, if $P'$ is the Pauli string resulting from $P$ by replacing every occurrence of $X$ and $Y$ by $Z$, $e^{-i \theta P'}$ applies the phase factor $e^{-i \theta}$ to an input computational basis state if its parity is even and $e^{i \theta}$ otherwise. The circuit for $e^{-i \theta P}$ can be obtained from that of $e^{-i \theta P'}$ by applying the suitable single-qubit basis transformation to the qubits where the respective Pauli operation in $P$ is $X$ or $Y$.

\begin{figure}[h]
\begin{tabular}{l}
(a) \raisebox{-0.8\totalheight}{
\begin{tikzpicture}
\node[scale=0.7]{
\begin{tikzcd}[row sep={0.55cm,between origins}, column sep=0.08cm]
\lstick{1} & \gate{H} & \qw & \ctrl{2} & \qw & \qw   & \qw       & \ctrl{2}      & \qw & \gate{H}  & \qw  \\
\lstick{2} & \qw      & \qw      & \qw      & \qw & \qw & \qw   & \qw      & \qw & \qw   & \qw \\
\lstick{3} & \gate{S^{\dagger}}      & \gate{H}      & \targ{}      & \ctrl{2} & \qw       & \ctrl{2} & \targ{} & \gate{H}  & \gate{S}       & \qw       \\
\lstick{4} & \qw  & \qw      & \qw      & \qw & \qw       & \qw       & \qw  & \qw      & \qw       & \qw       \\
\lstick{5} & \qw      & \qw  & \qw      & \targ{} & \gate{R_z(2\theta)}       & \targ{}       & \qw      & \qw      & \qw       & \qw       
\end{tikzcd}
};
\end{tikzpicture}
} \\
(b) \raisebox{-0.8\totalheight}{
\begin{tikzpicture}
\node[scale=0.7]{
\begin{tikzcd}[row sep={0.55cm,between origins}, column sep=0.08cm]
\lstick{1} & \gate{H} & \qw & \ctrl{1} & \targ{} & \ctrl{1} & \qw & \qw & \qw & \qw & \qw & \qw & \qw   & \qw       & \qw     & \qw & \qw  & \ctrl{1} & \targ{} & \ctrl{1} & \qw & \gate{H} & \qw \\
\lstick{2} & \qw      & \qw      & \targ{} & \ctrl{-1} & \targ{} & \ctrl{1} & \qw & \qw & \qw & \qw   & \qw & \qw & \qw   & \qw      & \qw & \ctrl{1}   & \targ{} & \ctrl{-1} & \targ{} & \qw & \qw & \qw \\
\lstick{3} & \gate{S^{\dagger}}      & \gate{H}      & \qw & \qw & \qw & \targ{}      & \ctrl{1} & \targ{} & \ctrl{1} & \qw  & \qw       & \qw & \ctrl{1} & \targ{}  & \ctrl{1}       & \targ{} & \qw & \qw & \qw & \gate{H} & \gate{S} & \qw      \\
\lstick{4} & \qw  & \qw     & \qw & \qw & \qw  & \qw     & \targ{} & \ctrl{-1} & \targ{} & \ctrl{1} & \qw & \ctrl{1}       & \targ{}       & \ctrl{-1}  & \targ{}      & \qw       & \qw  & \qw & \qw & \qw & \qw & \qw     \\
\lstick{5} & \qw      & \qw & \qw & \qw & \qw  & \qw      & \qw & \qw & \qw & \targ{} & \gate{R_z(2\theta)}       & \targ{}       & \qw      & \qw      & \qw       & \qw   & \qw & \qw & \qw & \qw & \qw & \qw       
\end{tikzcd}
};
\end{tikzpicture}
} \\
(c) \raisebox{-0.8\totalheight}{
\begin{tikzpicture}
\node[scale=0.7]{
\begin{tikzcd}[row sep={0.55cm,between origins}, column sep=0.08cm]
\lstick{1} & \gate{H} & \qw & \targ{} & \ctrl{1} & \qw & \qw & \qw & \qw & \qw & \qw   & \qw       & \qw     & \qw  & \ctrl{1} & \targ{} & \qw & \gate{H} & \qw \\
\lstick{2} & \qw      & \qw      & \ctrl{-1} & \targ{} & \ctrl{1} & \qw & \qw & \qw   & \qw & \qw & \qw   & \qw       & \ctrl{1}   & \targ{} & \ctrl{-1} & \qw & \qw & \qw \\
\lstick{3} & \gate{S^{\dagger}}      & \gate{H}      & \qw & \qw & \targ{}  & \targ{} & \ctrl{1} & \qw  & \qw       & \qw & \ctrl{1} & \targ{}         & \targ{} & \qw & \qw & \gate{H} & \gate{S} & \qw      \\
\lstick{4} & \qw  & \qw      & \qw & \qw  & \qw      & \ctrl{-1} & \targ{} & \ctrl{1} & \qw & \ctrl{1}       & \targ{}       & \ctrl{-1}      & \qw       & \qw  & \qw & \qw & \qw & \qw     \\
\lstick{5} & \qw      & \qw & \qw & \qw  & \qw       & \qw & \qw & \targ{} & \gate{R_z(2\theta)}       & \targ{}       & \qw      & \qw             & \qw   & \qw & \qw & \qw & \qw & \qw       
\end{tikzcd}
};
\end{tikzpicture}
}
\end{tabular}
\caption{Example of decomposition of quantum circuit that realizes $e^{-i \theta X_{1} \otimes Y_{3} \otimes Z_{5}}$ for $\theta \in \mathbb{R}$ for (a) all-to-all qubit connectivity, (b) linear qubit connectivity using \textsc{swap}s to reroute qubits, and (c) linear qubit connectivity using \textsc{cnot-swap}s to reroute qubits. The standard approach based on full \textsc{swap}s to move active qubits past idle ones yields a total of $16$ \textsc{cnot}s. The rerouting of the qubits with \textsc{cnot-swap}s saves $2$ \textsc{cnot}s for every idle qubit, thus resulting in an overall \textsc{cnot} count of $12$. Note also that the same strategy adopted in Fig.~\ref{fig:long-cnot}(c) to parallelize pairs of adjacent \textsc{cnot}s can also be employed here to reduce the circuit depth further.}
\label{fig:complex_exp_Pauli}
\end{figure}
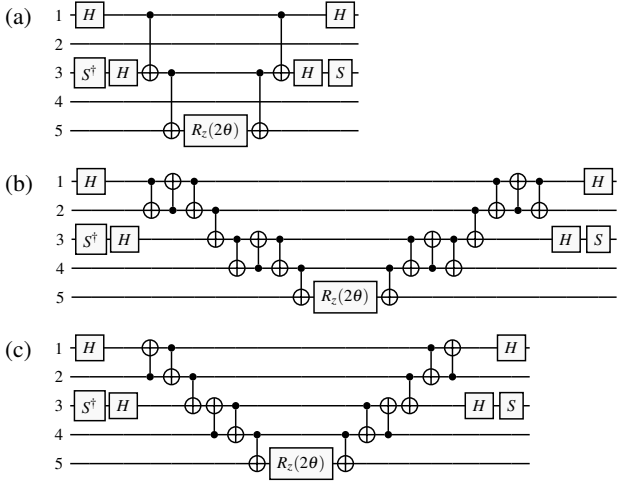

Under linear qubit connectivity, some of these $2(s-1)$ \textsc{cnot}s will be applied at pairs of non-adjacent qubits. The standard approach is to move one towards the other via \textsc{swap}s. However, once again every \textsc{swap} can be replaced by a \textsc{cnot-swap}, thus reducing the overall \textsc{cnot} count by $2$ for every idle qubit that is between the active qubits. Fig.~\ref{fig:complex_exp_Pauli} shows an example of a circuit for a complex exponential of a Pauli string, $e^{-i \theta X_{1} Y_{3} Z_{5}}$, under linear qubit connectivity; the \textsc{cnot} count obtained using \textsc{cnot-swap}s is $12$, i.e., with $4$ fewer \textsc{cnot} gates than the approach based on \textsc{swap} gates.

\section{Quantum circuit to prepare ground state of Fermi-Hubbard dimer via Gutzwiller wave function \label{AppD}}

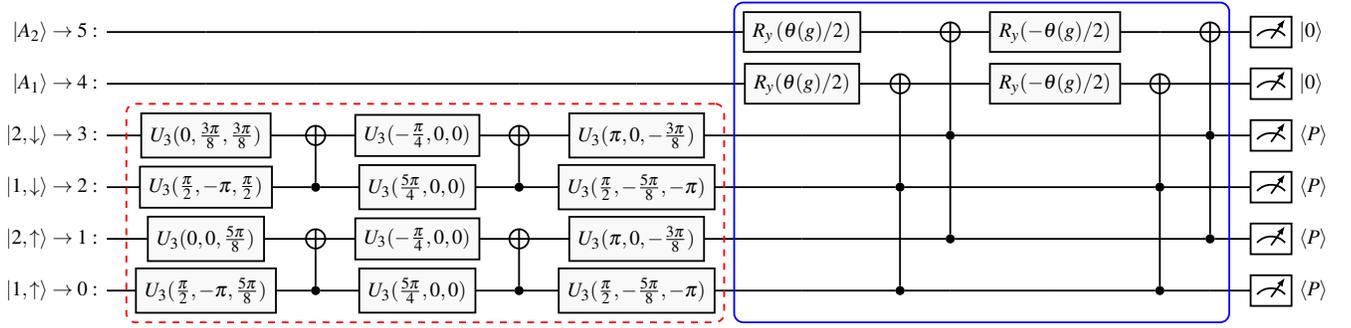
\begin{figure*}[t]
\flushleft
\begin{tikzpicture}
\node[scale=0.86]{
\begin{tikzcd}[column sep=0.45cm, row sep={0.8cm,between origins}]
\lstick{$\left| A_2 \right\rangle \rightarrow 5:$ } & \qw & \qw & \qw & \qw & \qw & \gate{R_y\left(\theta(g)/2\right)} \gategroup[6,steps=6,style={solid, rounded corners, draw=blue, inner xsep=1pt, inner ysep=1pt}]{} & \qw & \targ{} & \gate{R_y(-\theta(g)/2)} & \qw & \targ{} & \meter{}  \rstick{$\left|0\right\rangle$} \\
\lstick{$\left| A_1 \right\rangle \rightarrow 4:$ } & \qw & \qw & \qw & \qw & \qw & \gate{R_y(\theta(g)/2)} & \targ{} & \qw & \gate{R_y(-\theta(g)/2)} & \targ{} & \qw & \meter{}  \rstick{$\left|0\right\rangle$} \\
\lstick{$\left| 2,\downarrow \right\rangle \rightarrow 3:$ } & \gate{U_3(0,\frac{3\pi}{8},\frac{3\pi}{8})} \gategroup[4,steps=5,style={dashed, rounded corners, draw=red, inner xsep=1pt, inner ysep=01pt}]{} & \targ{} & \gate{U_3 (-\frac{\pi}{4},0,0)} & \targ{} & \gate{U_3(\pi,0,-\frac{3\pi}{8})} & \qw & \qw & \ctrl{-2} & \qw & \qw & \ctrl{-2} & \meter{} \rstick{ $\left \langle P \right \rangle$ } \\
\lstick{$\left|1,\downarrow \right\rangle \rightarrow 2:$ } & \gate{U_3(\frac{\pi}{2},-\pi,\frac{\pi}{2})} & \ctrl{-1} & \gate{U_3(\frac{5\pi}{4},0,0)} & \ctrl{-1} & \gate{U_3(\frac{\pi}{2},-\frac{5\pi}{8},-\pi)} & \qw & \ctrl{-2}  & \qw & \qw & \ctrl{-2} & \qw & \meter{} \rstick{ $\left \langle P \right \rangle$ } \\
\lstick{$\left| 2,\uparrow \right\rangle \rightarrow 1:$ } & \gate{U_3(0,0,\frac{5\pi}{8})} & \targ{} & \gate{U_3(-\frac{\pi}{4},0,0)} & \targ{} & \gate{U_3(\pi,0,-\frac{3\pi}{8})} & \qw & \qw & \ctrl{-2} & \qw & \qw & \ctrl{-2} & \meter{} \rstick{ $\left \langle P \right \rangle$ } \\
\lstick{$\left|1,\uparrow \right\rangle \rightarrow 0:$ } & \gate{U_3(\frac{\pi}{2},-\pi,\frac{5\pi}{8})} & \ctrl{-1} & \gate{U_3(\frac{5\pi}{4},0,0)} & \ctrl{-1} & \gate{U_3(\frac{\pi}{2},-\frac{5\pi}{8},-\pi)} & \qw & \ctrl{-2} & \qw & \qw & \ctrl{-2} & \qw & \meter{} \rstick{ $\left \langle P \right \rangle$ }
\end{tikzcd}
};
\end{tikzpicture}
\caption{Quantum circuit to prepare exact ground state of Fermi-Hubbard dimer via the Gutzwiller wave function\cite{murta2021gutzwiller} and compute its energy. For the special case of the dimer, the Gutzwiller wave function\cite{Gutzwiller1963} is the exact ground state of the Fermi-Hubbard model for $g = 1 - \frac{4t}{U + \sqrt{U^2 + 16t^2}}$, where $t$ is the hopping constant and $U$ is the Hubbard parameter. The first part of the circuit, shown inside the red dashed-line box, corresponds to the preparation of the ground state of the non-interacting model (i.e., for $\frac{U}{t} = 0$), which is just a Slater determinant\cite{Kivlichan2018, Jiang2018}. The corresponding subcircuit was decomposed in the $\{ U_3(\theta, \phi, \lambda), \textsc{cnot} \}$ basis to highlight the four \textsc{cnot}s. The second part of the circuit, shown inside the blue solid-line box, applies the Gutzwiller operator at each site non-deterministically, with $\theta(g) = 2 \arctan(\sqrt{2g - g^2}/(1-g))$. The preparation is successful when both ancillary qubits $A_1$ and $A_2$ are measured in the Z-basis and found in the $\ket{0}$ state. The success probability decreases with $\frac{U}{t}$, being $1$ for $\frac{U}{t} = 0$ and $\frac{1}{4}$ as $\frac{U}{t} \to \infty$. All-to-all qubit connectivity is assumed, so qubits do not need to be rerouted to perform the Toffoli gates, which require $6$ \textsc{cnot}s each. Once the ground state has been successfully prepared, its energy can be estimated by measuring all four qubits in the main register in the same single-qubit basis $P = X, Y, Z$, depending on the set of commuting terms --- $\{ X_0 X_1, X_2 X_3 \}$, $\{ Y_0 Y_1, Y_2 Y_3 \}$ or $\{ Z_{0}, Z_{1}, Z_{2}, Z_{3}, Z_{0} Z_{2}, Z_{1}Z_{3} \}$ --- that are computed. The qubit labels shown at the left end of the scheme are consistent with the expansion of the Hamiltonian of the Fermi-Hubbard dimer in the Pauli basis that is presented in Eq. (\ref{eq:Hamiltonian_sampling}) in the main text, assuming the Jordan-Wigner transformation to map electrons to qubits\cite{McArdle2020}.}
\label{fig:qc_FHM_gs}
\end{figure*}

The quantum circuit that was employed to prepare the ground state of the Fermi-Hubbard dimer in Section \ref{sec:ECA_Hubbard} is represented schematically in Fig.~\ref{fig:qc_FHM_gs}. It makes use of a quantum routine to prepare the Gutzwiller wave function \cite{murta2021gutzwiller} non-deterministically, which amounts to reducing the amplitude of basis states with doubly-occupied sites (see blue solid-line box) found in the non-interacting ground state (see red dashed-line box) to a degree set by the free parameter $g \in [0,1]$. For a given set of parameters $t$ and $U$ of the Fermi-Hubbard dimer, the Gutzwiller wave function is its exact ground state if $g = 1 - \frac{4t}{U + \sqrt{U^2 + 16t^2}}$. Fig.~\ref{fig:qc_FHM_gs} also details the order of the qubits that was assumed in the derivation of the expansion of the Hamiltonian in the Pauli basis (see Eq. (\ref{eq:Hamiltonian_sampling}) in the main text). The qubits were ordered by spin instead of site to avoid an extra $Z$ in the Pauli strings arising from the hopping terms. The four-qubit subcircuit inside the red dashed-line box, which prepares the non-interacting ground state, was decomposed in the $\{ U_{3}(\theta, \phi, \lambda), \textsc{cnot} \}$ basis, where $U_3(\theta, \phi, \lambda)$ is the general single-qubit operation
\begin{equation}
    U_3(\theta, \phi, \lambda) = \begin{pmatrix} 
    \cos(\theta/2) & -e^{i \lambda} \sin(\theta/2) \\
    e^{i\phi} \sin(\theta/2) & e^{i(\phi+\lambda)} \cos(\theta/2)
    \end{pmatrix}, 
    \label{eq:U-gate}
\end{equation}
in order to emphasize the $4$ \textsc{cnots} involved in its structure. Assuming all-to-all qubit connectivity, each of the four Toffoli gates in the subcircuit inside the blue solid-line box can be decomposed in terms of $6$ \textsc{cnot}s, resulting in a total \textsc{cnot} count of $28$.

\nocite{*}
\bibliography{biblio}

\end{document}